\documentclass[journal]{IEEEtran}

\usepackage[utf8]{inputenc}
\usepackage{graphics,graphicx,color,epsfig,subcaption}
\usepackage{amsmath,amssymb,amsthm,amsfonts,bm,bbm}
\usepackage{algorithmicx,algorithm,algpseudocode}
\algdef{SE}[DOWHILE]{Do}{doWhile}{\algorithmicdo}[1]{\algorithmicwhile\ #1}%
\usepackage{multirow}
\usepackage{cite}
\usepackage[normalem]{ulem}

\usepackage{tikz}
\usetikzlibrary{automata, positioning, arrows,shapes.multipart,fit}
\usepackage{pgfplots}

\newcommand{\todo}[1]{}
\newcommand{\added}[1]{#1}

\pgfplotsset{ignore legend/.style={every axis legend/.code={\renewcommand\addlegendentry[2][]{}}}}

\title{A Centralized Multi-stage Non-parametric Learning Algorithm for Opportunistic Spectrum Access}
\author{Thulasi Tholeti \hspace{16pt} Vishnu Raj \hspace{16pt} Sheetal Kalyani\\
    \hspace{-0 cm}Department of Electrical Engineering \\ Indian Institute of Technology, Madras \\
    Chennai, India 600 036 \\
    \texttt{\{ee15d410,ee14d213,skalyani\}@ee.iitm.ac.in}
}

\begin{document}
	\maketitle
	% \vspace{-2cm}
% 	\input{01_abstract.tex}
	
%     \input{10_intro.tex}
	
% 	\input{30_model.tex}
	
% 	\input{40_proposed.tex}
	
%   	\input{50_sim.tex}

% 	\input{60_conclusion.tex}
    \begin{abstract}
        Owing to the ever-increasing demand in  wireless spectrum, Cognitive Radio (CR) was introduced as a technique to attain high spectral efficiency. As the number of secondary users (SUs) connecting to the cognitive radio network is on the rise, there is an imminent need for centralized algorithms that provide high throughput and energy efficiency of the SUs while ensuring minimum interference to the licensed users. In this work, we propose a multi-stage algorithm that - 1) effectively assigns the available channel to the SUs, 2) employs a non-parametric learning framework to estimate the primary traffic distribution to minimize sensing, and 3) proposes an adaptive framework to ensure that the collision to the primary user is below the specified threshold. We provide comprehensive empirical validation of the method with other approaches.
    \end{abstract}
    
    
    % \section{Introduction}
    % Why cognitive radio?\\
    % Why is intelligent spectrum access important?\\
    % Typical scenarios where such a situation will arise\\
    % Previous work done\\
    % Motivation: Why learn? Why non-parametric\\
    % Our method highlights\\
    % Notation
    \section{Introduction}
Given the ever-increasing number of devices connecting to the network, the availability of spectrum is seen as the major bottleneck to achieve higher throughput. Methods to circumvent the purchase of spectrum or to improve spectral efficiency are gaining importance. In that context, Cognitive Radio is viewed as a potential solution to this problem \cite{mitola1999cognitive,khan2016cognitive,khan2017cognitive}. It enables the co-existence of licensed and unlicensed users in a spectrum aiming to improve the overall spectrum utilization in a wireless environment where spectrum resources are scarce \cite{liao2015full,sharma2015cognitive,tsiropoulos2016radio,ding2017cognitive}. The unlicensed users, commonly referred as Secondary Users (SUs), leverage holes available in the licensed spectrum, which are the result of spectrum under-utilization by Primary Users (PUs), to transmit their data. Since PUs own exclusive rights to access the allocated spectrum, SUs are required to ensure that the interference caused to the PU is minimal. This requires the SUs to sense the channel for presence of PU traffic when it wants to transmit. Each sensing operation comes with an associated cost of both energy and time spent on sensing the channels. There is always a trade-off between the throughput and interference caused to the PU.

Multiple approaches have been proposed for reducing the time spent by SUs on sensing the channels. Two of the popular approaches available in literature are to optimize (a) Channel Selection: the channels are ranked such that a channel which has a higher probability of being idle is selected first for sensing \cite{jiang2009optimal,canavitsas2013white,khan2013autonomous} and (b) Optimize inter-sensing interval: inter-sensing interval is calculated for each of the channels based on the available PU traffic statistics and the channels are sensed at these intervals instead of sensing the channel at the start of every frame \cite{pei2007sensing,oksanen2015order,senthilmurugan2017optimal}.

The channel selection problem in CRNs has been widely studied by formalizing it as a Reinforcement Learning (RL) problem. This includes posing it as a Multi-Armed Bandit (MAB) problem \cite{jouini2009multi,jouini2010upper,liu2010distributed,zhu2016multi}, applying Q-Learning \cite{van2011cooperative,hosey2009q} etc. Another popular approach followed for channel selection in CRN is combinatorial bandits \cite{gai2010learning,gai2011combinatorial} where each combination of channel allocation is seen as an action.
In \cite{bonnefoi2017multi}, a comparison study of different MAB algorithms is presented in the context of spectrum access in IoT networks. However, the channel has to be sensed before the transmission of every frame. 
% These methods do not leverage the fact that there are multiple SUs and the system can learn about the PU traffic by combining the sensing information from all the SUs and exploit the learned information to optimize the inter-sensing interval across on each channel.
The alternate approach to reduce the number of sensing operations required by the SUs and improve the system throughput is to optimize the inter-sensing interval by estimating the idle period of the PU. The work in \cite{pei2007sensing} proposes a framework for calculating the optimal frame duration for SUs to maximize the throughput while keeping the collision probability to PUs within a limit for an exponential traffic model. Later, \cite{senthilmurugan2017optimal} showed that PU traffic patterns can be best approximated with heavy tailed distributions and provided an optimal inter-sensing interval policy for HED traffic model. However, the parameters of primary channel were a requirement in the above works; this is not usually known to the SUs.

In \cite{raj2018spectrum}, a two-stage reinforcement learning method which combines the PU idle time estimation and channel ordering without the knowledge of channel parameters is proposed in a single SU scenario. By using a parametric Bayesian learning method to estimate the residual OFF time, the proposed scheme was able to learn an inter-sensing interval policy and combine it with a channel ordering policy based on MAB concepts.
% This work cannot be trivially extended to a multi-user scenario. Also, the work is model-based and uses an exponential model for Bayesian estimation. We are interested in extending this problem to a multi-user model-free solution. 
When multiple SUs are present in the CRN, the SUs may coordinate with each other or they may operate independently; each approach has its own pros and cons. In the CRN literature, works have adopted both the approaches according to the requirement of the CRN.

A cooperative spectrum sensing method where each SU experiences different channel environments is discussed in \cite{jin2019channel}. As an SU only senses the activity from its perception, a consensus between SUs on the PU state may not attained. The work proposes using a Hidden Markov Model (HMM) based technique considering results from all SUs to decide the state of the PU and further, shuts down the SUs which are likely to be affected by PU interference. In \cite{bhowmick2017throughput}, a cooperative method to predict and use the PU state to either exploit the spectrum hole (in case the PU is predicted to be inactive) or to harvest energy (in the case PU is predicted to be active) is proposed. In \cite{9056843}, a centralized algorithm to effectively combine the sensing decisions of the various nodes is proposed. A distributed spectrum sensing based on reinforcement learning is proposed in \cite{lunden2013multiagent} for multi-band sensing with multiple agents where the agents are assumed to be time-synchronized. In \cite{lunden2015distributed}, a study on sharing the spectrum sensing information in a distributed network using a common control channel is undertaken. Another solution to achieve coordination among the SUs for transmission in a distributed setup is proposed in \cite{zame2013cooperative}. This work achieves TDMA-like round-robin coordination among its SUs without any central supervision, only through distributed learning.
Interested readers are redirected to \cite{wang2016survey,wang2019survey} for a survey on learning-based techniques for more information.

Over the last decade, the number of devices that need to be connected such as smart phones, smart TVs, watches, sensors, etc. are on the rise. These numbers will further increase, especially with the introduction of commercial Internet of Things (IoT). All these devices may not be allocated a spectrum of their own as their throughput requirement may not be significant. Therefore, we now deal with a situation where a large number of secondary devices with arbitrary payload and throughput requirements will look for secondary access. Therefore, we propose to use a centralized framework where a central unit coordinates between the SUs not only to ensure that there are no inter-SU collisions but also to obtain samples from all the SUs to estimate the primary traffic behaviour.

In this paper, we propose a two-stage algorithm to effectively assign users to channels that are advantageous to them and once assigned, to skip the sensing operation based on the PU traffic estimated. In our method, we estimate both the probability of finding a primary channel idle as well as the time for which the channel is likely to stay idle once it is sensed free. The first goal is achieved through a modified search method known as hill climbing where the best possible channel-device pairing is identified. The second goal of PU traffic estimation is achieved through a model-free non-parametric Bayesian method without the need for apriori knowledge of the primary traffic parameters. \added{
In case a collision threshold is specified for the PU is, we also propose an adaptive method to balance transmission and sensing so as to exploit the threshold without causing excessive interference.} We compare the proposed method with others and draw inferences.

% In Section \ref{sec:sys_model}, the system model is described. The proposed approach is explained in Section \ref{sec:proposed} followed by the simulation results in Section \ref{sec:results}. We draw conclusions from our work and talk about the future scope in Section \ref{sec:Conc}.
    
    % \section{System Model}
    % Description of primary and secondary devices. State diagram only for SU. \\
    % Centralized vs distributed. Why centralized?\\
    % PU traffic models. Allowed collision threshold.\\
    % SU traffic models
    
    \section{System Model} \label{sec:sys_model}
    We consider a cognitive network where $\mathcal{N}$ denotes the set of PUs and $\mathcal{M}$ denotes the set of SUs, with $|\mathcal{N}| = N$ and $|\mathcal{M}| = M$. Each PU has its own licensed channel; there are $N$ channels available for secondary devices in the network for opportunistic access. 
    % The PUs and IoT devices are geographically separated in the scenario and hence the each IoT device will see different channel characteristics when using a particular band due to fading and propagation loss corruptions. 
    % For simplicity, we assume that the each PU will be in only one of the two states - either it will be transmitting data or it will be idle. Thus, 
    At any time, we have two sets of primary users, $\mathcal{N}_a$ denoting the set of active PUs and $\mathcal{N}_i$ denoting the set of idle PUs with $\mathcal{N}_a \cup \mathcal{N}_i = \mathcal{N}$ and $\mathcal{N}_a \cap \mathcal{N}_i = \phi$. The state transition diagram of PU is given in Figure \ref{fig:pu_state_trans}.

    \begin{figure}[!h]
        \centering
        \begin{tikzpicture}[scale=0.6,shorten >=1pt,>=stealth',node distance=4cm,on grid,every text node part/.style={align=center}]
            \node[state,minimum size=1.0cm] (pu_idle) {Idle\\$(PU \in \mathcal{N}_{i})$};
            \node[state,minimum size=1.0cm] (pu_active) [right=of pu_idle] {Active\\$(PU \in \mathcal{N}_{a})$};
            \path[->] (pu_idle) edge [bend right] node [below] {On data} (pu_active)
                      (pu_active) edge [bend right] node [above] {Data sent} (pu_idle)
                                  edge [loop above ] node   {On Collision} (); 
        \end{tikzpicture}
        \caption{PU's state transition diagram}
        \label{fig:pu_state_trans}
    \end{figure}

    % The primary user will stay in the idle state until some data needs to be transmitted. 
    When the PU has data to transmit, it moves from idle to active state and transmits on its designated channel. Hence PU stays in the active state until it successfully sends its data; even in the case that the PU encounters a collisions, it immediately resumes transmitting on its channel. Upon successful transmission, PU goes back to the idle state until the next set of data needs to be transmitted. We also assume that the PU allows a maximum threshold of a fraction of $T_{int}$ collision from the secondary users.

    % The case of IoT devices is different; they do not have any licensed bands and hence rely on opportunistic spectrum access for data transmission. This requires the IoT device to go through a sense-send cycle for data transmission. 
    
    % Whenever the IoT device requires to send any data, it will first have to wait until it gets assigned to a channel to sense. Upon sensing, if the channel is found to be free, the IoT device can use the channel to transmit data. 
    % The IoT device will move to idle state only after the data is successfully transmitted. We assume that the rate at which data is generated at the IoT device is low such that there will not be any queue buffer overflow at the device. This assumption is valid in various IoT scenarios where the payload will be of a few hundred bytes and the data updation interval will be of the order of minutes. 
    
    \begin{figure}[!h]
        \centering
        \begin{tikzpicture}[shorten >=1pt,>=stealth',node distance=4cm,on grid,every text node part/.style={align=center}]
            \node[state,minimum size=1.5cm] (iot_idle) {Idle\\$(D_j \in \mathcal{M}_{i})$};
            \node[state,minimum size=1.5cm] (iot_wait) [below left of=iot_idle] {Wait\\$(D_j \in \mathcal{M}_{w})$};
            \node[state,minimum size=1.5cm] (iot_active) [below right of=iot_idle] {Active\\$(D_j \in \mathcal{M}_{a})$};
            \node[state,minimum size=1.5cm] (iot_sense) [below right of=iot_wait] {Sense\\$(D_j \in \mathcal{M}_{s})$};
            {Active};
            \path[->] (iot_idle) edge [bend right] node [below] {On data} (iot_wait)
                      (iot_wait) edge [bend right] node [above] {Channel to sense} (iot_sense)
                      (iot_sense) edge [bend right] node [above] {Channel free} (iot_active)
                      (iot_sense) edge [bend right] node [below] {Channel busy} (iot_wait)
                      (iot_active) edge  node [above] {Collision} (iot_wait)
                      (iot_active) edge [bend right] node [below] {Data sent} (iot_idle);
        \end{tikzpicture}
        \caption{SU device $D_j$'s transition diagram}
        \label{fig:iot_state_trans}
    \end{figure}

    In our work, we consider a centralized framework for the SUs where a central node $U$ is responsible for coordinating with the secondary devices. There are four disjoint sets of SUs, idle devices denoted by $\mathcal{M}_i$, devices waiting for channel access denoted by $\mathcal{M}_w$, devices in channel sensing phase denoted by $\mathcal{M}_s$ and devices which are active (transmitting data) denoted by $\mathcal{M}_a$. At every iteration $t$, $U$ checks for SUs in the wait state ($|\mathcal{M}_{w,t}| > 0$). If a device $D_j$ is in the wait state, $U$ assigns one of the channels, $\mathcal{C}_{j,t}$, to that device to sense. The device $D_j$ senses the channel $\mathcal{C}_{j,t}$ and reports the observation back to $U$. If the channel is not free, the SU will return to the wait state, and wait until it is given another channel to sense. If the channel $C_{j,t}$ is sensed to be free, the SU sensing it can access the channel to transmit its data. Upon successful transmission, the SU moves to idle state. If the transmission is unsuccessful, the device returns to the wait state having achieved zero throughput. The state transition diagram of an SU $D_j$ is given in Figure \ref{fig:iot_state_trans}.
     \todo{Frame structure and synchronization.}

    For primary user traffic we consider two continuous time traffic models based on the recent empirical studies \cite{raj2018spectrum}: Generalized Pareto Distributed (GPD) model and  Exponential model. 
    \begin{enumerate}
        \item \textbf{Exponential Model:} Both ON and OFF times of the primary traffic are modelled as exponential distributions with parameters $\lambda_1$ and $\lambda_2$. These parameters can be varied to modify the intensity of traffic. This model is adopted by other works in the CRN literature such as \cite{pei2007sensing,senthilmurugan2017optimal} and \cite{raj2018spectrum}.
        \item \textbf{Generalized Pareto Model}: Both the ON time and OFF time of PU is distributed as Generalized Pareto Distribution (GPD). The probability density function of GPD is given by
        \begin{align}
            f_{X}^{GPD}(x|k,\sigma,\theta) = \frac{1}{\sigma} \left( 1 + k \frac{x-\theta}{\sigma} \right)^{-1-\frac{1}{k}},
        \end{align}
        where $x > \theta$ and $k > 0$. Here $k, \sigma$ and $\theta$ are shape, scale and location parameters respectively. 
        % Mean of the distribution is given by $ \mu  + \frac{\sigma}{1-k}$ for $k < 1$. 
        Different traffic characteristics are captured by varying the value of parameters. For example, the percentage occupancy in a band by PU can be modelled by varying the location parameter of the ON and OFF distributions. GPD is also shown to approximate real traffic in \cite{stabellini2010quantifying}.
        % \item \textbf{Hyper Exponential Model}: HED traffic model is based on the observation that PUs will have long OFF periods with short ON periods. The idle time characteristics of the PU have been captured by this distribution in \cite{stabellini2010quantifying,senthilmurugan2017optimal}. To capture this behaviour, HED model uses Exponential distribution to model ON time and HED distribution to model OFF times. Thus the ON time distribution of HED model with mean ON time as $\mu_{ON}$ is given by
        % \begin{align}
        %     f_{X}^{HED-ON}(x|\mu) = \frac{1}{\mu_{ON}} \exp \left( -\frac{x}{\mu_{ON}}\right)
        % \end{align}
        % and OFF time distribution with mean OFF period $\sum \frac{p_i}{\mu_i}$ is given by
        % \begin{align}
        %     f_{X}^{HED-OFF}(x|\bar{p},\bar{\mu}) = \sum p_i \frac{1}{\mu_{i}} \exp \left( -\frac{x}{\mu_{i}}\right).
        % \end{align}
    \end{enumerate}
    % For simulation studies, we chose the parameters of the models to closely resemble the empirical observations which reflect real life PU traffic use cases. 
    % It should also be noted that Exponential traffic can be generated as a special case of Hyper ExD by changing the OFF time distribution to have only one component with $p_1 = 1$.
    We model the SU traffic as periodic traffic where a fixed number of frames is transmitted periodically by the device. This type of traffic is one of the commonly encountered IoT traffic models \cite{nikaein2013simple}. It can also be used for various sensor devices which report their readings periodically. 

    \section{Proposed approach} \label{sec:proposed}
We propose a centralized approach for spectrum sensing where a central node communicates with all the secondary devices. Note that as the devices are not ON permanently, they only observe a part of the primary spectrum. This limited number of samples will not allow any single device to formulate a good estimate of the PU traffic. Therefore, we adopt the centralized approach over the distributed as our aim is to use samples from all the devices. In addition to this advantage, using a centralized approach is also beneficial as all the computation is done at the central node and hence, energy-constrained SUs can operate effectively. The role of the central node is to collect data from various SUs, thereby having more information to form a better estimate of the PU behaviour than any individual device. The central unit then uses this estimate to predict which channel is more likely to be available and the duration for which the channel would be available for secondary transmission. \\
Our proposed algorithm operates has two functionalities:
\begin{enumerate}
    \item Channel assignment - The central node assigns channels to devices in \(\mathcal{M}_w\) for sensing.
    \item Residual time prediction - Once a channel is assigned to an SU, it is likely to be available for more than one just frame length. This module predicts for which the PU stays idle on the chosen channels.
\end{enumerate}
The interactions between the SU Node, Central Unit and the channel is presented in the schematic in Figure \ref{fig:schematic}. We shall elaborate on these modules in the rest of the section.
    % \begin{figure*}[h]
    %     \centering
    %     \includegraphics[scale = 0.9]{JIOT/Lilliput_map.pdf}
    %     % \input{figs/comm_diagram.tex}
    %     \caption{Interactions in the CRN}
    %     \label{fig:schematic}
    % \end{figure*}
    
    \begin{figure}[h]
        \centering
        \tikzstyle{block} = [rectangle, draw, node distance=6cm, text width=6em, text centered, rounded corners, minimum height=2em]
\tikzstyle{event} = [draw=none, fill=white, circle, minimum size=0.001cm]
\tikzstyle{pathlabel} = [above, sloped]

\begin{tikzpicture}[auto,>=stealth',every text node part/.style={align=center},scale=0.7, every node/.style={transform shape}]
    \begin{scope}[auto,node distance=1.0cm]
        \node [block] (hub) {Central Hub};
        \node [block, right of = hub, node distance = 4cm] (su) {SU};
        \node [block, right of = su, node distance = 4cm] (channel) {Channel};
        
        \node [event, below of = su, node distance = 1cm] (req_chlist_start){};
        % \filldraw (req_chlist_start) circle(0.4mm);
        \node [event, below of = req_chlist_start, xshift = -4cm] (req_chlist_end){};
        % \filldraw (req_chlist_end) circle(0.4mm);
        \draw[->,black] (req_chlist_start) -- node[pathlabel] {$\mathcal{M}_w$} (req_chlist_end);
        
        \node [event, below of = req_chlist_end, node distance = 0.10cm] (send_chlist_start){};
        % \filldraw (send_chlist_start) circle(0.4mm);
        \node [event, below of = send_chlist_start, xshift = +4cm] (send_chlist_end){};
        % \filldraw (send_chlist_end) circle(0.4mm);
        \draw[->,black] (send_chlist_start) -- node[pathlabel] {Channel List} (send_chlist_end);
        
        \node [event, below of = send_chlist_end, node distance = 0.10cm] (sense_start){};
        % \filldraw (sense_start) circle(0.4mm);
        \node [event, right of = sense_start, xshift = +3cm] (sense_end){};
        % \filldraw (sense_end) circle(0.4mm);
        \draw[<->,black] (sense_start) -- node[pathlabel] {Sense} (sense_end);
        
        \node [event, below of = sense_start, node distance = 0.10cm] (sense_result_start){};
        % \filldraw (sense_result_start) circle(0.4mm);
        \node [event, below of = sense_result_start, xshift = -4cm] (sense_result_end){};
        % \filldraw (sense_result_end) circle(0.4mm);
        \draw[->,black] (sense_result_start) -- node[pathlabel] {Sense Result} (sense_result_end);
        
        \node [event, below of = sense_result_end, node distance = 0.10cm] (ch_select_start){};
        % \filldraw (ch_select_start) circle(0.4mm);
        \node [event, below of = ch_select_start, xshift = 4cm] (ch_select_end){};
        % \filldraw (ch_select_end) circle(0.4mm);
        \draw[->,black] (ch_select_start) -- node[pathlabel] {$c, t_{skip}$} (ch_select_end);
        
        \node [event, below of = ch_select_end, node distance = 0.10cm] (tx1_start){};
        % \filldraw (tx1_start) circle(0.4mm);
        \node [event, below of = tx1_start, xshift = 4cm] (tx1_end){};
        % \filldraw (tx1_end) circle(0.4mm);
        \draw[->,black] (tx1_start) -- node[pathlabel] {Frame 1} (tx1_end);
        
        \node [event, below of = tx1_start, node distance = 0.90cm] (tx2_start){};
        % \filldraw (tx2_start) circle(0.4mm);
        \node [event, below of = tx2_start, xshift = 4cm] (tx2_end){};
        % \filldraw (tx2_end) circle(0.4mm);
        \draw[->,black] (tx2_start) -- node[pathlabel] {Frame 2} (tx2_end);
        
        \node [event, below of = tx2_start, node distance = 0.90cm] (tx3_start){};
        % \filldraw (tx3_start) circle(0.4mm);
        \node [event, below of = tx3_start, xshift = 4cm] (tx3_end){};
        % \filldraw (tx3_end) circle(0.4mm);
        \draw[->,black] (tx3_start) -- node[pathlabel](p1) {Frame 2} (tx3_end);
        
        \node [event, below of = tx3_start, node distance = 2.00cm] (txN_start){};
        % \filldraw (txN_start) circle(0.4mm);
        \node [event, below of = txN_start, xshift = 4cm] (txN_end){};
        % \filldraw (txN_end) circle(0.4mm);
        \draw[->,black] (txN_start) -- node[pathlabel](p2) {Frame $N$} (txN_end);
        
        \draw[dotted] (p1) -- (p2);
        
        \node [event, left of = txN_end, node distance = 0.10cm, xshift = -4cm] (feedback_start){};
        % \filldraw (feedback_start) circle(0.4mm);
        \node [event, below of = feedback_start, xshift = -4cm] (feedback_end){};
        % \filldraw (feedback_end) circle(0.4mm);
        \draw[->,black] (feedback_start) -- node[pathlabel] {Throughput,\\ Time Spent} (feedback_end);
        
        \node [below of = feedback_end, node distance = 0.50cm] (hub_end) {};
        \filldraw (hub_end) circle(0.4mm);
        \node [right of = hub_end, node distance = 4cm] (su_end) {};
        \filldraw (su_end) circle(0.4mm);
        \node [right of = su_end, node distance = 4cm] (channel_end) {};
        \filldraw (channel_end) circle(0.4mm);
        
        \draw[] (hub) -- (hub_end);
        \draw[] (su) -- (su_end);
        \draw[] (channel) -- (channel_end);
        
    \end{scope}
\end{tikzpicture}
        \caption{Time line of interactions in the CRN with proposed approach}
        \label{fig:schematic}
    \end{figure}

\subsection{Channel Assignment}
Different secondary nodes can achieve different throughput on the same channel. The aim of this module is two-fold: firstly, the central node learns the characteristics of primary traffic to predict the availability and secondly, the central node also assigns the channels to those devices so as to maximize the sum throughput. Note that the central node does not have any prior information regarding the PU traffic and should learn the same from scratch.

In the case of single SU, channel selection using MAB is quite popular\cite{jouini2009multi,zhu2016multi}. However, we deal with multiple SUs that demand for a channel at the same instant. This problem reduces to assigning the best user-channel permutation in the case that we know the value of each user-channel pairing. However, we do not have access to that value and hence learn that from data. A similar problem is dealt in \cite{gai2011combinatorial} using combinatorial bandits; however, their solution is restricted to the case where the number of channels is greater than the number of users, both of which do not change with time. In our formulation, the number of active users and the number of available channels change with time. Hence, we need to search over all possible permutations to arrive at a channel assignment. This is very computationally demanding. For example, then we have 5 free channels and 10 SUs requesting for channels, the search space is $10!/5! = 30240$. Therefore we propose to use a technique called hill climbing which has substantially low complexity.
    
% We employ a learning technique which combines the ideas of AI method, hill climbing \cite{russel178artificial}, and reinforcement learning technique called $\epsilon$-greedy \cite{auer2002using}. 
The algorithm proceeds by estimating a value for each of the channel-device pairs; the value table is stored at the central node. We represent each entry of this table by $V_{c,d}$. Whenever a feedback on throughput, $\mathcal{T}$, is available from the device, the corresponding entry in the value table is updated according to the update equation
    \begin{align}
        V_{c,d} \gets \kappa \cdot \mathcal{T}  + (1-\kappa) \cdot V_{c,d}.
    \end{align}
Here $\kappa$ is a problem dependent parameter, also known as \textit{learning rate}. When we set $\kappa = 1$, the central hub gives importance to only last observation and completely discards any of the past learning. Conversely, if $\kappa$ is very close to $0$, the central hub will take long time to build up the value table as it give very less weight to new observations.
    
With the value table representing the estimate of quality of each channel for each device, we can calculate the quality of each channel assignment configuration based on the individual entries in the table. The total value can be calculated as the sum of the individual values of the configuration. One obvious way of channel assignment is to search over all possible configurations with the available value estimates. To avoid the computational burden of the brute force search, we propose using the hill climbing algorithm to arrive at a good configuration \cite{russell2016artificial}. Hill climbing proceeds by randomly swapping two pairs of entries in the channel-device assignment and recalculating the quality of resultant configuration. If the new configuration has a higher total value than the last configuration, we discard the previous configuration and use the new one to proceed. This process is continued until no swaps improve the total value of the channel assignment configuration.
    
Note that the value table maintained at the central node needs to be estimated correctly for hill climbing to work and the values are learnt from scratch. Therefore, the algorithm will tend to be biased to the devices that it initially picks. To combat this, we need an exploration-exploitation strategy. Hence, we employ the $\epsilon$-greedy strategy to randomly explore different configurations with a small probability. By trying random configurations $\eta$ fraction of the time, the central node can improve the accuracy of the value table over time. This, in turn, makes the results of hill climbing better. The methods for channel assignment in central node is provided in Algorithm \ref{alg:hub_channel}. \todo{Mention this happens at every frame? Mention input and output. Explain swap mathematically. }

\begin{algorithm}
        \caption{Central Node - Sub routines for channel assignment}
        \label{alg:hub_channel}
        \begin{algorithmic}[1]
            % \Function{GetChannel}{$\mathcal{M}_w$}
                \State \textbf{Initialization:} Channel assignment configuration, $\mathcal{Z}_0$;
                $V_{c,d} = 0 \quad \forall c,d$;
                Iteration counter $j=0$.
                % \State Initialize channel assignment configuration with random $\mathcal{Z}_0$
                % \State Compute total value $v_0$ for $\mathcal{Z}_0$ as from the current value table
                % \State Set counter j = 0
                \Do
                    \State Create a random swap of $\mathcal{Z}_j$ to get $\hat{\mathcal{Z}}_{j}$
                    \State Calculate total value $\hat{f}_j$ for $\hat{\mathcal{Z}}_{j}$
                    \If{ $\hat{f}_j \geq f_j$}
                        \State $\mathcal{Z}_{j+1} = \hat{\mathcal{Z}_j}$ 
                        \State $j \gets j + 1$
                    \EndIf
                    % \State Exit when no new swap results in improved total value
                \doWhile{Any new swap results in improved total value}
                % \For{Each device $j \in \mathcal{M}_w$}
                %     \State With probability $1-\eta$, assign channel $c^*$ to device $j$ such that $c^* = \underset{c \in \mathcal{U}_t}{\arg \max} \: Q_{c,j} $ 
                %     \State With probability $\eta$, assign a random channel $c^* \in  \mathcal{U}_t$ to device $j$
                %     \State $\mathcal{U}_t \gets \mathcal{U}_t \setminus c^*$
                % \EndFor
                \State With probability $1-\eta$, use the channel allocation configuration $\mathcal{Z}_{j}$
                \State With probability $\eta$, do random channel assignment configuration $\mathcal{Z}_0$
            % \EndFunction
            % \Function{UpdateChannel}{$d$,$c$,$\mathcal{T}$}
                % \If{transmission successful}
                %     \State $S_c \gets S_c + 1$
                % \Else
                %     \State $F_c \gets F_c + 1$
                % \EndIf
                \State Employ configuration to transmit and obtain the throughput $\mathcal{T}$ for every pair
                \State For every pair $V_{c,d} \gets \kappa \cdot \mathcal{T} + (1-\kappa) \cdot V_{c,d} $
            % \EndFunction
        \end{algorithmic}
    \end{algorithm}
    
    \subsection{Residual Time Prediction}
    This block, present in the central node, comes into play once the channel assignment is done. It estimates the amount of time for which the PU is likely to stay idle after the sensing operation; in other words, it estimates the residual OFF time online by observing the idle times that are reported by all the SUs.  This in turn is used to gauge the number of frames for which sensing can be skipped and the SU can transmit its data on the assigned channel without sensing. We propose a Bayesian method of estimation using the past samples of residual OFF times observed by all the SUs. Though we are estimating a continuous quantity - the residual OFF times, we only have access to samples quantized to the frame length. Also, we need to predict the number of frames that can be transmitted. We therefore construct a discrete distribution from the observed samples and sample from the estimated distribution to predict the residual OFF time of an assigned channel. Note that although a similar Bayesian estimation method is employed in \cite{raj2018spectrum}, it treats the residual OFF time as a continuous quantity and employs a exponential model whereas the proposed method is model-free.
    
    \begin{algorithm}
        \caption{Residual time prediction for a channel $c$}
        \label{alg:hub_residual}
        \begin{algorithmic}[1]
            \State \textbf{Initialization} Create a Dirichlet distribution $\mathcal{D}_c$ with parameters $a_{c,1},\ldots,a_{c,\bar{K}}$
            \For{$k = 1,2, \cdots$}
                \State Sample a multinomial distribution $\textbf{p} \sim \mathcal{D}_{c}$
                \State Create an augmented distribution $\hat{\textbf{p}} = (1-\epsilon_t) \cdot \textbf{p} + \epsilon_t \cdot \delta(\bar{K})$
                \State Sample from augmented distribution, $t_{skip} \sim \hat{\textbf{p}}$
                \State Instruct SU to skip for $t_{skip}$ frames 
                \State Obtain feedback of number of frames successfully skipped $\tau$
                \State Update $\mathcal{D}_{c}$ as $a_{c,\tau} \gets a_{c,\tau} + 1$ 
            \EndFor
        \end{algorithmic}
    \end{algorithm}
    
    With the problem of estimating the residual OFF time reduced to estimating a discrete distribution, we aim to build a categorical distribution to estimate the number of frames to be skipped. We assume a support of $\{0,1,\cdots, \bar{K}\}$, where $\bar{K}$ is a large number as we are not aware of the highest number of frames to skip. As the number of classes in the categorical distribution that is to be estimated is not known previously, we term our method as a non-parametric estimation method. We denote the categorical distribution of the residual OFF times quantized to frame length as $T_{R} \sim Cat(\bar{K},\textbf{p})$ with $\bar{K}$ classes and the probability of the occurrence of each class is denoted by the vector $\textbf{p}$. A Dirichlet distribution is used as the prior as it is the conjugate prior. The Dirichlet distribution, $\textbf{p} \sim Dir(\bar{K},\textbf{a})$ is parameterized by $\textbf{a} = (a_1, \cdots a_{\bar{K}})$ with $a_i > 0$ for $1 \leq i \leq \bar{K}$. The support of $\mathcal{D}_{\bar{K}}$ is a $(\bar{K}-1)$-dimensional simplex $S_{\bar{K}}$. The probability density function of $\textbf{p} = (p_1,\ldots,p_{\bar{K}})$ when $\textbf{p} \in S_{\bar{K}}$ is given by
    \begin{align*}
        D(p_1,\ldots,p_{\bar{K}};a_1,\ldots,a_{\bar{K}})
            = \frac{\Gamma\left(\sum \limits_{i=1}^{\bar{K}} a_i\right)}
                    {\prod \limits_{i=1}^{\bar{K}} \Gamma(a_i) }
                        \prod \limits_{i=1}^{\bar{K}} p_i^{a_i-1}.
    \end{align*}
    
    Let us assume that we obtain samples $(\tau_1, \cdots, \tau_{\bar{k}})$ where $\tau_i$ refers to the number of occurrences corresponding to class $i$. Note that this sample ($\tau_i$) is the number of successful frame transmissions without sensing and can fall in any of the 3 categories: SU skips sensing for the number of frames suggested by the central node, SU transmits its entire payload successfully or SU encounters a collision. Then, the posterior distribution is a Dirichlet distribution with the parameters updated as follows
    \begin{align*}
        a_i \gets a_i + \tau_i \quad \forall i= \{0,1,\cdots, \bar{K}\}.
    \end{align*}
    Note that this update can be performed online as the 
central node receives feedback samples from each of the SUs. When performed online, a obtain one sample at a time and the Dirichlet parameter corresponding to that class is incremented by one. When an assigned channel is sensed free, a probability vector $\textbf{p}$ is sampled from the posterior Dirichlet distribution and a sample from the categorical distribution is picked according to the obtained probability vector $\textbf{p}$. This result is then communicated to the SU as the number of frames for which sensing can be skipped, which we denote as $t_{skip}$.

Summarizing, the central node is responsible for predicting the residual OFF time for each of the channels. For predicting the residual OFF time of channel $c$, we first sample a categorical distribution with parameter $\textbf{p}$ from the maintained Dirichlet prior and then sample a point , $t_{skip}$, from $\textbf{p}$, which corresponds to the discrete quantized time to skip. This is sent to the SUs to indicate the number of frames it can send without sensing. By sampling from the prior distribution and then sampling from the categorical $\textbf{p}$, one ensures that with non-zero probability the central node will try to explore various skip periods. Since the central node is building the distribution by also taking actions based on the past observed values of residual OFF time, it needs to try transmit for longer times than what it has already observed to build the tail part of posterior distributions. The functionality of the Residual Time Prediction module is described in Algorithm \ref{alg:hub_residual}.

    \noindent
    \textbf{Prior for the Dirichlet distribution:} Note that as the number of classes of the Dirichlet distribution, a large number \(\bar{K}\) is chosen. Therefore, while setting the prior distribution, it is not advisable to use a uniform distribution over the entire support. To effectively set the prior for each channel, we initially transmit until a collision is observed. Let us call the number of frames successfully transmitted as \(N_{init}\). The aim is to set the concentration parameters \(\alpha_i\)'s such that the Dirichlet prior concentrates around \(N_{init}/2\). To achieve that, we set
    \begin{align*}
        \alpha_1 = \begin{cases}
            i              &; i = 1, \cdots, N_{init}/2,\\
            \alpha_{i-1}/2 &; i = N_{init}/2, \cdots, \bar{K}.
        \end{cases}
    \end{align*}
    This focuses the prior around a region that is more likely to be a sample of the OFF period. Note that this prior is set in an adhoc fashion and there may be other methods of setting the prior.
    
    \noindent
    \textbf{Exploring higher residual OFF times:}
    Rather than using only the Bayesian sampling technique to explore various residual OFF periods, we can make the central node explicitly try high values of skip periods to build the tail of the categorical distribution. To do so, we sample according to the maintained prior for $1-\epsilon$ fraction of the time and for $\epsilon$ fraction of the time, we set $t_{skip}$ as $\bar{K}$ which is a high number. By scaling $\textbf{p}$ with $1-\epsilon$ and adding a mass of $\epsilon$ at $\bar{K}$, the exploration is incorporated into the Bayesian method. We denote this augmented distribution by $\hat{\textbf{p}}$. Using $\hat{\textbf{p}}$ instead of $\textbf{p}$ will cause higher collision. Hence, the exploration factor $\epsilon$ can be selected such that the experienced collisions is within the allowed threshold of interference, $T_{int}$. We wish to adaptively learn the parameter $\epsilon$ from the observed behaviour to ensure maximum efficiency - allowed interference trade-off.
    \begin{algorithm}   
        \caption{Adaptively modify $\epsilon$ for a given channel}    \label{alg:adaptive_exploration}
        \begin{algorithmic}[1]
            \State \textbf{Initialization:} Set $k =1$, $count =1$, $a$, $v$, $\alpha$, $\gamma$ and $\epsilon_0$
            \For{$count = 1,2, \cdots$}
                \State $a_k= \dfrac{a}{k^\alpha}$
                \State $v_k= \dfrac{v}{k^\gamma}$
                \State $d = sign(g - T_{int})$
                \If{$count$ is odd}
                    \State Set $\epsilon_t =  \epsilon_k $ and observe $L(\epsilon_k)$
                \Else
                    \State Set $\epsilon_t =  \epsilon_k - d v_k$ and observe $L(\epsilon_k - d v_k)$
                    \State $\hat{g} = \dfrac{L(\epsilon_k) - L(\epsilon_k -d v_k)}{ v_k }$
                    \State $\epsilon_{k+1} = \epsilon_k - d a_k \hat{g}$
                    \State $k = k+1$
                \EndIf
            \EndFor
        \end{algorithmic}
    \end{algorithm}

    \added{We know that the percentage of frame collisions, say $g$, is affected by $\epsilon$ and other factors. Let us denote it as the function $g(\epsilon,\bar{\theta})$ where $\bar{\theta}$ denotes the other factors. From the SU's perspective, it is ideal if $g(\epsilon,\bar{\theta})$ is very close to (just below) $T_{int}$ as it achieves the efficiency vs interference trade-off. Also, note that when $g(\epsilon,\bar{\theta})> T_{int}$, the exploration must be drastically reduced but when $g(\epsilon,\bar{\theta})< T_{int}$, $\epsilon$ may be gradually increased. For convenience, we write denote the percentage of observed interference as $g$. Therefore, we propose the following asymmetric loss function \cite{parthasarathy2018interference}:
    \begin{equation} \label{eqn:asymmetric}
       L(\epsilon) =  l(g, T_{int}) = \begin{cases}
        \exp{|g-T_{int}|} -1, & g > T_{int},\\
        (g-T_{int})^2, & g < T_{int}.
        \end{cases}
    \end{equation}
    As the quantities $g$ and $T_{int}$ are in the range $[0,1]$, the exponential always has a higher slope than the quadratic function. Note that we do not know the mathematical form of $g(\epsilon,\bar{\theta})$ and hence need to estimate the gradient by sampling the loss function. Inspired by Finite Difference Stochastic Averaging (FDSA) method \cite{bhatnagar2012stochastic}, we propose Algorithm \ref{alg:adaptive_exploration} to adaptively modify $\epsilon$. }
    
    Here, $k$ denotes the number of updates performed on the channel $c$ whereas $count$ denotes the iterations where the central node is asked to predict number of frames to skip on a specific channel. The parameters $a,\alpha,v,\gamma$ are problem dependent and need to be tuned. By adapting the exploration to the PU traffic seen, we can ensure that we maximally utilize the spectrum. In the following section, we present the simulation studies for the proposed approach.
    
    % A main research problem in reinforcement learning is addressing how to control the exploratory behaviour of the agent without losing the ability to learn. Below, we discuss three different ways to choose $\epsilon$.
    % One of the most popular methods is to keep $\epsilon$ a constant so that the SU explores with a constant probability.
    % \begin{align}
    %     \epsilon_{t+1} \gets \epsilon.   \label{eqn:hub_explore_const}
    % \end{align}
    
    % One of the main disadvantage of constant exploration is that the cumulative penalty associated with exploratory actions will increase linearly over time; an undesired characteristic for any learning algorithm. If we can appropriately decay the exploration factor over time, then we can counter this linearly increasing cumulative regret. Exponentially decaying the exploration factor with time is also a popular approach \cite{auer2002using}. 
    % \begin{align}
    %     \epsilon_{t+1} \gets \frac{1}{t^\beta}. \label{eqn:hub_explore_decay}
    % \end{align}
    % A high value of $\beta$ can lead to sub-optimal exploration whereas a low value can lead to very slow learning process.
    % The optimal value of decaying parameter $\beta$ is problem dependent. We now see if we can adaptively calculate the exploration factor based on the observed PU traffic behaviour.

    \section{Results} \label{sec:results}
\ifCLASSOPTIONtwocolumn
\begin{figure*}[t]
    \centering
    \begin{subfigure}{.33\textwidth}
        \resizebox{\linewidth}{!}{
            \pgfplotstableread[col sep = comma]{./data_new/data_Exp100050_FC.csv}\datatable
            \pgfplotsset{ylabel=Avg. frame collision, ymin=-0.05, ymax=+0.25, ytick={-0.05,0.0,...,+0.30}, ignore legend}
            \tikzstyle{plot_style} = [line width=1.25pt, mark size={4.0}, mark repeat=4, mark phase=1]
\begin{tikzpicture}[thick]
    \begin{axis}[
        width=8cm,
        height=7cm,
        xmin=0e3,
        xmax=14e3,
        grid=major,
        xlabel={Timesteps},
        xlabel style={font=\large},
        ylabel style={font=\large},
        ytick style={font=\tiny},
        yticklabel style={
            /pgf/number format/.cd,
                fixed,
                fixed zerofill,
                precision=2,
            /tikz/.cd
        },
        % log ticks with fixed point,
        % title={},
        % legend pos=
        legend style={at={(0.99,0.40)},anchor=south east},
        legend cell align={left},
        legend style={fill opacity=0.80, draw opacity=0.50, text opacity=1.0}
        ]
        
        % \addplot[black, solid, thick, mark=+, plot_style] 
        %     table [y=Genetic_TUC_y, x=Genetic_TUC_x, col sep=comma]{\datatable};
        % \addlegendentry{TUC};
        \addplot[black, solid, thick, mark=+, plot_style] 
            table [y=Random_TUC_y, x=Random_TUC_x, col sep=comma]{\datatable};
        \addlegendentry{TUC};
        
        % \addplot[black, solid, thick, mark=x, plot_style] 
        %     table [y=Genetic_Traditional_y, x=Genetic_Traditional_x, col sep=comma]{\datatable};
        % \addlegendentry{Traditional};
        \addplot[black, solid, thick, mark=x, plot_style] 
            table [y=Random_Traditional_y, x=Random_Traditional_x, col sep=comma]{\datatable};
        \addlegendentry{Traditional};
        
        % \addplot[black, solid, thick, mark=triangle, plot_style] 
        %     table [y=Genetic_Raj2018_y, x=Genetic_Raj2018_x, col sep=comma]{\datatable};
        % \addlegendentry{Augmented \cite{raj2018spectrum}};
        \addplot[black, solid, thick, mark=diamond, plot_style] 
            table [y=Random_Raj2018_y, x=Random_Raj2018_x, col sep=comma]{\datatable};
        \addlegendentry{\cite{raj2018spectrum}};
        
        % \addplot[black, solid, thick, mark=diamond, plot_style] 
        %     table [y=Genetic_Proposed_Fixed_eps_0_y, x=Genetic_Proposed_Fixed_eps_0_x, col sep=comma]{\datatable};
        % \addlegendentry{Proposed Greedy};
        
        \addplot[black, solid, thick, mark=square, plot_style] 
            table [y=Genetic_Proposed_Fixed_eps_0_1_y, x=Genetic_Proposed_Fixed_eps_0_1_x, col sep=comma]{\datatable};
        \addlegendentry{Proposed $(\epsilon = 0.10)$};
        
        % \addplot[black, solid, thick, mark=pentagon, plot_style] 
        %     table [y=Genetic_Proposed_Decreasing_eps_y, x=Genetic_Proposed_Decreasing_eps_x, col sep=comma]{\datatable};
        % \addlegendentry{Proposed ($\epsilon \downarrow$)};
        
        \addplot[black, solid, thick, mark=o, plot_style] 
            table [y=Genetic_Proposed_FDSA_SqExp_0_13_y, x=Genetic_Proposed_FDSA_SqExp_0_13_x, col sep=comma]{\datatable};
        \addlegendentry{Proposed FDSA $(T_{int} = 0.13)$};
    \end{axis}
\end{tikzpicture}
        }
        \caption{Avg Frame Collision}
        \label{fig:data_Exp100050_C05_S10PE_FC}
    \end{subfigure}%
    \begin{subfigure}{.33\textwidth}
        \resizebox{\linewidth}{!}{
            \pgfplotstableread[col sep = comma]{./data_new/data_Exp100050_TP.csv}\datatable
            \pgfplotsset{ylabel=Avg. Throughput, ymin=2.00, ymax=3.20, ytick={2.00,2.20,...,3.30}, ignore legend}
            \tikzstyle{plot_style} = [line width=1.25pt, mark size={4.0}, mark repeat=4, mark phase=1]
\begin{tikzpicture}[thick]
    \begin{axis}[
        width=8cm,
        height=7cm,
        xmin=0e3,
        xmax=14e3,
        grid=major,
        xlabel={Timesteps},
        xlabel style={font=\large},
        ylabel style={font=\large},
        ytick style={font=\tiny},
        yticklabel style={
            /pgf/number format/.cd,
                fixed,
                fixed zerofill,
                precision=2,
            /tikz/.cd
        },
        % log ticks with fixed point,
        % title={},
        % legend pos=
        legend style={at={(0.99,0.40)},anchor=south east},
        legend cell align={left},
        legend style={fill opacity=0.80, draw opacity=0.50, text opacity=1.0}
        ]
        
        % \addplot[black, solid, thick, mark=+, plot_style] 
        %     table [y=Genetic_TUC_y, x=Genetic_TUC_x, col sep=comma]{\datatable};
        % \addlegendentry{TUC};
        \addplot[black, solid, thick, mark=+, plot_style] 
            table [y=Random_TUC_y, x=Random_TUC_x, col sep=comma]{\datatable};
        \addlegendentry{TUC};
        
        % \addplot[black, solid, thick, mark=x, plot_style] 
        %     table [y=Genetic_Traditional_y, x=Genetic_Traditional_x, col sep=comma]{\datatable};
        % \addlegendentry{Traditional};
        \addplot[black, solid, thick, mark=x, plot_style] 
            table [y=Random_Traditional_y, x=Random_Traditional_x, col sep=comma]{\datatable};
        \addlegendentry{Traditional};
        
        % \addplot[black, solid, thick, mark=triangle, plot_style] 
        %     table [y=Genetic_Raj2018_y, x=Genetic_Raj2018_x, col sep=comma]{\datatable};
        % \addlegendentry{Augmented \cite{raj2018spectrum}};
        \addplot[black, solid, thick, mark=diamond, plot_style] 
            table [y=Random_Raj2018_y, x=Random_Raj2018_x, col sep=comma]{\datatable};
        \addlegendentry{\cite{raj2018spectrum}};
        
        % \addplot[black, solid, thick, mark=diamond, plot_style] 
        %     table [y=Genetic_Proposed_Fixed_eps_0_y, x=Genetic_Proposed_Fixed_eps_0_x, col sep=comma]{\datatable};
        % \addlegendentry{Proposed Greedy};
        
        \addplot[black, solid, thick, mark=square, plot_style] 
            table [y=Genetic_Proposed_Fixed_eps_0_1_y, x=Genetic_Proposed_Fixed_eps_0_1_x, col sep=comma]{\datatable};
        \addlegendentry{Proposed $(\epsilon = 0.10)$};
        
        % \addplot[black, solid, thick, mark=pentagon, plot_style] 
        %     table [y=Genetic_Proposed_Decreasing_eps_y, x=Genetic_Proposed_Decreasing_eps_x, col sep=comma]{\datatable};
        % \addlegendentry{Proposed ($\epsilon \downarrow$)};
        
        \addplot[black, solid, thick, mark=o, plot_style] 
            table [y=Genetic_Proposed_FDSA_SqExp_0_13_y, x=Genetic_Proposed_FDSA_SqExp_0_13_x, col sep=comma]{\datatable};
        \addlegendentry{Proposed FDSA $(T_{int} = 0.13)$};
    \end{axis}
\end{tikzpicture}
        }
        \caption{Achievable Throughput}
        \label{fig:data_Exp100050_C05_S10PE_TP}
    \end{subfigure}%
    \begin{subfigure}{.33\textwidth}
        \resizebox{\linewidth}{!}{
            \pgfplotstableread[col sep = comma]{./data_new/data_Exp100050_SE.csv}\datatable
            \pgfplotsset{ylabel=Avg. sensing, ymin=0.80,ymax=2.20, ytick={0.80,1.00,...,2.30}}
            \tikzstyle{plot_style} = [line width=1.25pt, mark size={4.0}, mark repeat=4, mark phase=1]
\begin{tikzpicture}[thick]
    \begin{axis}[
        width=8cm,
        height=7cm,
        xmin=0e3,
        xmax=14e3,
        grid=major,
        xlabel={Timesteps},
        xlabel style={font=\large},
        ylabel style={font=\large},
        ytick style={font=\tiny},
        yticklabel style={
            /pgf/number format/.cd,
                fixed,
                fixed zerofill,
                precision=2,
            /tikz/.cd
        },
        % log ticks with fixed point,
        % title={},
        % legend pos=
        legend style={at={(0.99,0.40)},anchor=south east},
        legend cell align={left},
        legend style={fill opacity=0.80, draw opacity=0.50, text opacity=1.0}
        ]
        
        % \addplot[black, solid, thick, mark=+, plot_style] 
        %     table [y=Genetic_TUC_y, x=Genetic_TUC_x, col sep=comma]{\datatable};
        % \addlegendentry{TUC};
        \addplot[black, solid, thick, mark=+, plot_style] 
            table [y=Random_TUC_y, x=Random_TUC_x, col sep=comma]{\datatable};
        \addlegendentry{TUC};
        
        % \addplot[black, solid, thick, mark=x, plot_style] 
        %     table [y=Genetic_Traditional_y, x=Genetic_Traditional_x, col sep=comma]{\datatable};
        % \addlegendentry{Traditional};
        \addplot[black, solid, thick, mark=x, plot_style] 
            table [y=Random_Traditional_y, x=Random_Traditional_x, col sep=comma]{\datatable};
        \addlegendentry{Traditional};
        
        % \addplot[black, solid, thick, mark=triangle, plot_style] 
        %     table [y=Genetic_Raj2018_y, x=Genetic_Raj2018_x, col sep=comma]{\datatable};
        % \addlegendentry{Augmented \cite{raj2018spectrum}};
        \addplot[black, solid, thick, mark=diamond, plot_style] 
            table [y=Random_Raj2018_y, x=Random_Raj2018_x, col sep=comma]{\datatable};
        \addlegendentry{\cite{raj2018spectrum}};
        
        % \addplot[black, solid, thick, mark=diamond, plot_style] 
        %     table [y=Genetic_Proposed_Fixed_eps_0_y, x=Genetic_Proposed_Fixed_eps_0_x, col sep=comma]{\datatable};
        % \addlegendentry{Proposed Greedy};
        
        \addplot[black, solid, thick, mark=square, plot_style] 
            table [y=Genetic_Proposed_Fixed_eps_0_1_y, x=Genetic_Proposed_Fixed_eps_0_1_x, col sep=comma]{\datatable};
        \addlegendentry{Proposed $(\epsilon = 0.10)$};
        
        % \addplot[black, solid, thick, mark=pentagon, plot_style] 
        %     table [y=Genetic_Proposed_Decreasing_eps_y, x=Genetic_Proposed_Decreasing_eps_x, col sep=comma]{\datatable};
        % \addlegendentry{Proposed ($\epsilon \downarrow$)};
        
        \addplot[black, solid, thick, mark=o, plot_style] 
            table [y=Genetic_Proposed_FDSA_SqExp_0_13_y, x=Genetic_Proposed_FDSA_SqExp_0_13_x, col sep=comma]{\datatable};
        \addlegendentry{Proposed FDSA $(T_{int} = 0.13)$};
    \end{axis}
\end{tikzpicture}
        }
        \caption{Avg No.of Sensing per active SU}
        \label{fig:data_Exp100050_C05_S10PE_SE}
    \end{subfigure}%
    \caption{Results for Exponential Model for $5$ channels and $10$ users for periodic SU traffic}
    \label{fig:Exp_comp}
\end{figure*}

\begin{figure*}[t]
    \centering
    \begin{subfigure}{.33\textwidth}
        \resizebox{\linewidth}{!}{
            \pgfplotstableread[col sep = comma]{./data_new/data_Exp100050_FC.csv}\datatable
            \pgfplotsset{ylabel=Avg. frame collision, ymin=0.11, ymax=0.17, ytick={0.11,0.12,...,0.18}, ignore legend}
            \tikzstyle{plot_style} = [line width=1.25pt, mark size={4.0}, mark repeat=4, mark phase=1]
\begin{tikzpicture}[thick]
    \begin{axis}[
        width=8cm,
        height=7cm,
        xmin=0e3,
        xmax=14e3,
        grid=major,
        xlabel={Timesteps},
        xlabel style={font=\large},
        ylabel style={font=\large},
        ytick style={font=\tiny},
        yticklabel style={
            /pgf/number format/.cd,
                fixed,
                fixed zerofill,
                precision=2,
            /tikz/.cd
        },
        % log ticks with fixed point,
        % title={},
        legend pos=north east,
        legend cell align={left},
        legend style={fill opacity=0.80, draw opacity=0.50, text opacity=1.0}
        ]
        
        \addplot[black, solid, thick, mark=triangle, plot_style] 
            table [y=Genetic_Proposed_FDSA_SqExp_0_11_y, x=Genetic_Proposed_FDSA_SqExp_0_11_x, col sep=comma]{\datatable};
        \addlegendentry{$T_{int} = 0.11$};
        
        \addplot[black, solid, thick, mark=square, plot_style] 
            table [y=Genetic_Proposed_FDSA_SqExp_0_13_y, x=Genetic_Proposed_FDSA_SqExp_0_13_x, col sep=comma]{\datatable};
        \addlegendentry{$T_{int} = 0.13$};
        
        \addplot[black, solid, thick, mark=pentagon, plot_style] 
            table [y=Genetic_Proposed_FDSA_SqExp_0_15_y, x=Genetic_Proposed_FDSA_SqExp_0_15_x, col sep=comma]{\datatable};
        \addlegendentry{$T_{int} = 0.15$};
        
        \addplot[black, solid, thick, mark=o, plot_style] 
            table [y=Genetic_Proposed_FDSA_SqExp_0_18_y, x=Genetic_Proposed_FDSA_SqExp_0_18_x, col sep=comma]{\datatable};
        \addlegendentry{$T_{int} = 0.18$};
    \end{axis}
\end{tikzpicture}
        }
        \caption{Avg Frame Collision}
        \label{fig:data_Exp100050_C05_S10PE_FC_TAU}
    \end{subfigure}%
    \begin{subfigure}{.33\textwidth}
        \resizebox{\linewidth}{!}{
            \pgfplotstableread[col sep = comma]{./data_new/data_Exp100050_TP.csv}\datatable
            \pgfplotsset{ylabel=Avg. Throughput, ymin=2.70, ymax=3.10, ytick={2.70, 2.80,...,3.20}, ignore legend}
            \tikzstyle{plot_style} = [line width=1.25pt, mark size={4.0}, mark repeat=4, mark phase=1]
\begin{tikzpicture}[thick]
    \begin{axis}[
        width=8cm,
        height=7cm,
        xmin=0e3,
        xmax=14e3,
        grid=major,
        xlabel={Timesteps},
        xlabel style={font=\large},
        ylabel style={font=\large},
        ytick style={font=\tiny},
        yticklabel style={
            /pgf/number format/.cd,
                fixed,
                fixed zerofill,
                precision=2,
            /tikz/.cd
        },
        % log ticks with fixed point,
        % title={},
        legend pos=north east,
        legend cell align={left},
        legend style={fill opacity=0.80, draw opacity=0.50, text opacity=1.0}
        ]
        
        \addplot[black, solid, thick, mark=triangle, plot_style] 
            table [y=Genetic_Proposed_FDSA_SqExp_0_11_y, x=Genetic_Proposed_FDSA_SqExp_0_11_x, col sep=comma]{\datatable};
        \addlegendentry{$T_{int} = 0.11$};
        
        \addplot[black, solid, thick, mark=square, plot_style] 
            table [y=Genetic_Proposed_FDSA_SqExp_0_13_y, x=Genetic_Proposed_FDSA_SqExp_0_13_x, col sep=comma]{\datatable};
        \addlegendentry{$T_{int} = 0.13$};
        
        \addplot[black, solid, thick, mark=pentagon, plot_style] 
            table [y=Genetic_Proposed_FDSA_SqExp_0_15_y, x=Genetic_Proposed_FDSA_SqExp_0_15_x, col sep=comma]{\datatable};
        \addlegendentry{$T_{int} = 0.15$};
        
        \addplot[black, solid, thick, mark=o, plot_style] 
            table [y=Genetic_Proposed_FDSA_SqExp_0_18_y, x=Genetic_Proposed_FDSA_SqExp_0_18_x, col sep=comma]{\datatable};
        \addlegendentry{$T_{int} = 0.18$};
    \end{axis}
\end{tikzpicture}
        }
        \caption{Achievable Throughput}
        \label{fig:data_Exp100050_C05_S10PE_TP_TAU}
    \end{subfigure}%
    \begin{subfigure}{.33\textwidth}
        \resizebox{\linewidth}{!}{
            \pgfplotstableread[col sep = comma]{./data_new/data_Exp100050_SE.csv}\datatable
            \pgfplotsset{ylabel=Avg. sensing, ymin=0.80, ymax=1.40, ytick={0.90,1.00,...,1.30}}
            \tikzstyle{plot_style} = [line width=1.25pt, mark size={4.0}, mark repeat=4, mark phase=1]
\begin{tikzpicture}[thick]
    \begin{axis}[
        width=8cm,
        height=7cm,
        xmin=0e3,
        xmax=14e3,
        grid=major,
        xlabel={Timesteps},
        xlabel style={font=\large},
        ylabel style={font=\large},
        ytick style={font=\tiny},
        yticklabel style={
            /pgf/number format/.cd,
                fixed,
                fixed zerofill,
                precision=2,
            /tikz/.cd
        },
        % log ticks with fixed point,
        % title={},
        legend pos=north east,
        legend cell align={left},
        legend style={fill opacity=0.80, draw opacity=0.50, text opacity=1.0}
        ]
        
        \addplot[black, solid, thick, mark=triangle, plot_style] 
            table [y=Genetic_Proposed_FDSA_SqExp_0_11_y, x=Genetic_Proposed_FDSA_SqExp_0_11_x, col sep=comma]{\datatable};
        \addlegendentry{$T_{int} = 0.11$};
        
        \addplot[black, solid, thick, mark=square, plot_style] 
            table [y=Genetic_Proposed_FDSA_SqExp_0_13_y, x=Genetic_Proposed_FDSA_SqExp_0_13_x, col sep=comma]{\datatable};
        \addlegendentry{$T_{int} = 0.13$};
        
        \addplot[black, solid, thick, mark=pentagon, plot_style] 
            table [y=Genetic_Proposed_FDSA_SqExp_0_15_y, x=Genetic_Proposed_FDSA_SqExp_0_15_x, col sep=comma]{\datatable};
        \addlegendentry{$T_{int} = 0.15$};
        
        \addplot[black, solid, thick, mark=o, plot_style] 
            table [y=Genetic_Proposed_FDSA_SqExp_0_18_y, x=Genetic_Proposed_FDSA_SqExp_0_18_x, col sep=comma]{\datatable};
        \addlegendentry{$T_{int} = 0.18$};
    \end{axis}
\end{tikzpicture}
        }
        \caption{Avg No.of Sensing per active SU}
        \label{fig:data_Exp100050_C05_S10PE_SE_TAU}
    \end{subfigure}%
    \caption{Effect of $T_{int}$ for Exponential Model for $5$ channels and $10$ users using Proposed FDSA}
    \label{fig:exp_Tint}
\end{figure*}
\fi

\ifCLASSOPTIONtwocolumn
\begin{figure*}[t]
    \centering
    \begin{subfigure}{.33\textwidth}
        \resizebox{\linewidth}{!}{
            \pgfplotstableread[col sep = comma]{./data_new/data_GPD025_FC.csv}\datatable
            \pgfplotsset{ylabel=Avg. Frame collision, ymin=-0.05, ymax=+0.20, ytick={-0.05,0.00,...,+0.25}, ignore legend}
            \tikzstyle{plot_style} = [line width=1.25pt, mark size={4.0}, mark repeat=4, mark phase=1]
\begin{tikzpicture}[thick]
    \begin{axis}[
        width=8cm,
        height=7cm,
        xmin=0e3,
        xmax=12e3,
        grid=major,
        xlabel={Timesteps},
        xlabel style={font=\large},
        ylabel style={font=\large},
        ytick style={font=\tiny},
        yticklabel style={
            /pgf/number format/.cd,
                fixed,
                fixed zerofill,
                precision=2,
            /tikz/.cd
        },
        % log ticks with fixed point,
        % title={},
        % legend pos=south east,
        legend style={at={(0.99,0.40)},anchor=south east},
        legend cell align={left},
        legend style={fill opacity=0.80, draw opacity=0.50, text opacity=1.0}
        ]
        
        % \addplot[black, solid, thick, mark=+, plot_style] 
        %     table [y=Genetic_TUC_y, x=Genetic_TUC_x, col sep=comma]{\datatable};
        % \addlegendentry{TUC};
        \addplot[black, solid, thick, mark=+, plot_style] 
            table [y=Random_TUC_y, x=Random_TUC_x, col sep=comma]{\datatable};
        \addlegendentry{TUC};
        
        % \addplot[black, solid, thick, mark=x, plot_style] 
        %     table [y=Genetic_Traditional_y, x=Genetic_Traditional_x, col sep=comma]{\datatable};
        % \addlegendentry{Traditional};
        \addplot[black, solid, thick, mark=x, plot_style] 
            table [y=Random_Traditional_y, x=Random_Traditional_x, col sep=comma]{\datatable};
        \addlegendentry{Traditional};
        
        % \addplot[black, solid, thick, mark=triangle, plot_style] 
        %     table [y=Genetic_Raj2018_y, x=Genetic_Raj2018_x, col sep=comma]{\datatable};
        % \addlegendentry{Augmented \cite{raj2018spectrum}};
        \addplot[black, solid, thick, mark=diamond, plot_style] 
            table [y=Random_Raj2018_y, x=Random_Raj2018_x, col sep=comma]{\datatable};
        \addlegendentry{\cite{raj2018spectrum}};
        
        % \addplot[blue, solid, thick, mark=diamond, plot_style] 
        %     table [y=Random2_Raj2018_y, x=Random2_Raj2018_x, col sep=comma]{\datatable};
        % \addlegendentry{Adhoc2 \cite{raj2018spectrum}}
        
        % \addplot[black, solid, thick, mark=diamond, plot_style] 
        %     table [y=Genetic_Proposed_Fixed_eps_0_y, x=Genetic_Proposed_Fixed_eps_0_x, col sep=comma]{\datatable};
        % \addlegendentry{Proposed Greedy};
        
        \addplot[black, solid, thick, mark=square, plot_style] 
            table [y=Genetic_Proposed_Fixed_eps_0_15_y, x=Genetic_Proposed_Fixed_eps_0_15_x, col sep=comma]{\datatable};
        \addlegendentry{Proposed $(\epsilon = 0.15)$};
        
        % \addplot[black, solid, thick, mark=pentagon, plot_style] 
        %     table [y=Genetic_Proposed_Decreasing_eps_y, x=Genetic_Proposed_Decreasing_eps_x, col sep=comma]{\datatable};
        % \addlegendentry{Proposed ($\epsilon \downarrow$)};
        
        \addplot[black, solid, thick, mark=o, plot_style] 
            table [y=Genetic_Proposed_FDSA_SqExp_0_13_y, x=Genetic_Proposed_FDSA_SqExp_0_13_x, col sep=comma]{\datatable};
        \addlegendentry{Proposed FDSA $(T_{int}=0.13)$};
    \end{axis}
\end{tikzpicture}
        }
        \caption{Avg Frame Collision}
        \label{fig:data_GPD025_C05_S10PE_FC}
    \end{subfigure}%
    \begin{subfigure}{.33\textwidth}
        \resizebox{\linewidth}{!}{
            \pgfplotstableread[col sep = comma]{./data_new/data_GPD025_TP.csv}\datatable
            \pgfplotsset{ylabel=Avg. Throughput, ymin=2.5, ymax=4.50, ytick={2.50,3.00,...,4.50}, ignore legend}
            \tikzstyle{plot_style} = [line width=1.25pt, mark size={4.0}, mark repeat=4, mark phase=1]
\begin{tikzpicture}[thick]
    \begin{axis}[
        width=8cm,
        height=7cm,
        xmin=0e3,
        xmax=12e3,
        grid=major,
        xlabel={Timesteps},
        xlabel style={font=\large},
        ylabel style={font=\large},
        ytick style={font=\tiny},
        yticklabel style={
            /pgf/number format/.cd,
                fixed,
                fixed zerofill,
                precision=2,
            /tikz/.cd
        },
        % log ticks with fixed point,
        % title={},
        % legend pos=south east,
        legend style={at={(0.99,0.40)},anchor=south east},
        legend cell align={left},
        legend style={fill opacity=0.80, draw opacity=0.50, text opacity=1.0}
        ]
        
        % \addplot[black, solid, thick, mark=+, plot_style] 
        %     table [y=Genetic_TUC_y, x=Genetic_TUC_x, col sep=comma]{\datatable};
        % \addlegendentry{TUC};
        \addplot[black, solid, thick, mark=+, plot_style] 
            table [y=Random_TUC_y, x=Random_TUC_x, col sep=comma]{\datatable};
        \addlegendentry{TUC};
        
        % \addplot[black, solid, thick, mark=x, plot_style] 
        %     table [y=Genetic_Traditional_y, x=Genetic_Traditional_x, col sep=comma]{\datatable};
        % \addlegendentry{Traditional};
        \addplot[black, solid, thick, mark=x, plot_style] 
            table [y=Random_Traditional_y, x=Random_Traditional_x, col sep=comma]{\datatable};
        \addlegendentry{Traditional};
        
        % \addplot[black, solid, thick, mark=triangle, plot_style] 
        %     table [y=Genetic_Raj2018_y, x=Genetic_Raj2018_x, col sep=comma]{\datatable};
        % \addlegendentry{Augmented \cite{raj2018spectrum}};
        \addplot[black, solid, thick, mark=diamond, plot_style] 
            table [y=Random_Raj2018_y, x=Random_Raj2018_x, col sep=comma]{\datatable};
        \addlegendentry{\cite{raj2018spectrum}};
        
        % \addplot[blue, solid, thick, mark=diamond, plot_style] 
        %     table [y=Random2_Raj2018_y, x=Random2_Raj2018_x, col sep=comma]{\datatable};
        % \addlegendentry{Adhoc2 \cite{raj2018spectrum}}
        
        % \addplot[black, solid, thick, mark=diamond, plot_style] 
        %     table [y=Genetic_Proposed_Fixed_eps_0_y, x=Genetic_Proposed_Fixed_eps_0_x, col sep=comma]{\datatable};
        % \addlegendentry{Proposed Greedy};
        
        \addplot[black, solid, thick, mark=square, plot_style] 
            table [y=Genetic_Proposed_Fixed_eps_0_15_y, x=Genetic_Proposed_Fixed_eps_0_15_x, col sep=comma]{\datatable};
        \addlegendentry{Proposed $(\epsilon = 0.15)$};
        
        % \addplot[black, solid, thick, mark=pentagon, plot_style] 
        %     table [y=Genetic_Proposed_Decreasing_eps_y, x=Genetic_Proposed_Decreasing_eps_x, col sep=comma]{\datatable};
        % \addlegendentry{Proposed ($\epsilon \downarrow$)};
        
        \addplot[black, solid, thick, mark=o, plot_style] 
            table [y=Genetic_Proposed_FDSA_SqExp_0_13_y, x=Genetic_Proposed_FDSA_SqExp_0_13_x, col sep=comma]{\datatable};
        \addlegendentry{Proposed FDSA $(T_{int}=0.13)$};
    \end{axis}
\end{tikzpicture}
        }
        \caption{Achievable Throughput}
        \label{fig:data_GPD025_C05_S10PE_TP}
    \end{subfigure}%
    \begin{subfigure}{.33\textwidth}
        \resizebox{\linewidth}{!}{
            \pgfplotstableread[col sep = comma]{./data_new/data_GPD025_SE.csv}\datatable
            \pgfplotsset{ylabel=Avg. sensing, ymin=0.60, ymax=1.80, ytick={0.60,0.80,...,1.80}}
            \tikzstyle{plot_style} = [line width=1.25pt, mark size={4.0}, mark repeat=4, mark phase=1]
\begin{tikzpicture}[thick]
    \begin{axis}[
        width=8cm,
        height=7cm,
        xmin=0e3,
        xmax=12e3,
        grid=major,
        xlabel={Timesteps},
        xlabel style={font=\large},
        ylabel style={font=\large},
        ytick style={font=\tiny},
        yticklabel style={
            /pgf/number format/.cd,
                fixed,
                fixed zerofill,
                precision=2,
            /tikz/.cd
        },
        % log ticks with fixed point,
        % title={},
        % legend pos=south east,
        legend style={at={(0.99,0.40)},anchor=south east},
        legend cell align={left},
        legend style={fill opacity=0.80, draw opacity=0.50, text opacity=1.0}
        ]
        
        % \addplot[black, solid, thick, mark=+, plot_style] 
        %     table [y=Genetic_TUC_y, x=Genetic_TUC_x, col sep=comma]{\datatable};
        % \addlegendentry{TUC};
        \addplot[black, solid, thick, mark=+, plot_style] 
            table [y=Random_TUC_y, x=Random_TUC_x, col sep=comma]{\datatable};
        \addlegendentry{TUC};
        
        % \addplot[black, solid, thick, mark=x, plot_style] 
        %     table [y=Genetic_Traditional_y, x=Genetic_Traditional_x, col sep=comma]{\datatable};
        % \addlegendentry{Traditional};
        \addplot[black, solid, thick, mark=x, plot_style] 
            table [y=Random_Traditional_y, x=Random_Traditional_x, col sep=comma]{\datatable};
        \addlegendentry{Traditional};
        
        % \addplot[black, solid, thick, mark=triangle, plot_style] 
        %     table [y=Genetic_Raj2018_y, x=Genetic_Raj2018_x, col sep=comma]{\datatable};
        % \addlegendentry{Augmented \cite{raj2018spectrum}};
        \addplot[black, solid, thick, mark=diamond, plot_style] 
            table [y=Random_Raj2018_y, x=Random_Raj2018_x, col sep=comma]{\datatable};
        \addlegendentry{\cite{raj2018spectrum}};
        
        % \addplot[blue, solid, thick, mark=diamond, plot_style] 
        %     table [y=Random2_Raj2018_y, x=Random2_Raj2018_x, col sep=comma]{\datatable};
        % \addlegendentry{Adhoc2 \cite{raj2018spectrum}}
        
        % \addplot[black, solid, thick, mark=diamond, plot_style] 
        %     table [y=Genetic_Proposed_Fixed_eps_0_y, x=Genetic_Proposed_Fixed_eps_0_x, col sep=comma]{\datatable};
        % \addlegendentry{Proposed Greedy};
        
        \addplot[black, solid, thick, mark=square, plot_style] 
            table [y=Genetic_Proposed_Fixed_eps_0_15_y, x=Genetic_Proposed_Fixed_eps_0_15_x, col sep=comma]{\datatable};
        \addlegendentry{Proposed $(\epsilon = 0.15)$};
        
        % \addplot[black, solid, thick, mark=pentagon, plot_style] 
        %     table [y=Genetic_Proposed_Decreasing_eps_y, x=Genetic_Proposed_Decreasing_eps_x, col sep=comma]{\datatable};
        % \addlegendentry{Proposed ($\epsilon \downarrow$)};
        
        \addplot[black, solid, thick, mark=o, plot_style] 
            table [y=Genetic_Proposed_FDSA_SqExp_0_13_y, x=Genetic_Proposed_FDSA_SqExp_0_13_x, col sep=comma]{\datatable};
        \addlegendentry{Proposed FDSA $(T_{int}=0.13)$};
    \end{axis}
\end{tikzpicture}
        }
        \caption{Avg No.of Sensing per active SU}
        \label{fig:data_GPD025_C05_S10PE_SE}
    \end{subfigure}%
    \caption{Results for GPD Model for $5$ channels and $10$ users for periodic SU traffic}
    \label{fig:GPD_comp}
\end{figure*}

\begin{figure*}[t]
    \centering
    \begin{subfigure}{.33\textwidth}
        \resizebox{\linewidth}{!}{
            \pgfplotstableread[col sep = comma]{./data_new/data_GPD025_FC.csv}\datatable
            \pgfplotsset{ylabel=Avg. Frame collision, ymin=0.09, ymax=0.14, ignore legend}
            \tikzstyle{plot_style} = [line width=1.25pt, mark size={4.0}, mark repeat=4, mark phase=1]
\begin{tikzpicture}[thick]
    \begin{axis}[
        width=8cm,
        height=7cm,
        xmin=0e3,
        xmax=12e3,
        grid=major,
        xlabel={Timesteps},
        xlabel style={font=\large},
        ylabel style={font=\large},
        ytick style={font=\tiny},
        yticklabel style={
            /pgf/number format/.cd,
                fixed,
                fixed zerofill,
                precision=2,
            /tikz/.cd
        },
        % log ticks with fixed point,
        % title={},
        legend pos=north east,
        legend cell align={left},
        legend style={fill opacity=0.80, draw opacity=0.50, text opacity=1.0}
        ]
        
        \addplot[black, solid, thick, mark=+, plot_style] 
            table [y=Genetic_Proposed_FDSA_SqExp_0_10_y, x=Genetic_Proposed_FDSA_SqExp_0_10_x, col sep=comma]{\datatable};
        \addlegendentry{$T_{int} = 0.10$};
        
        % \addplot[black, solid, thick, mark=triangle, plot_style] 
        %     table [y=Genetic_Proposed_FDSA_SqExp_0_11_y, x=Genetic_Proposed_FDSA_SqExp_0_11_x, col sep=comma]{\datatable};
        % \addlegendentry{$T_{int} = 0.11$};
        
        \addplot[black, solid, thick, mark=square, plot_style] 
            table [y=Genetic_Proposed_FDSA_SqExp_0_12_y, x=Genetic_Proposed_FDSA_SqExp_0_12_x, col sep=comma]{\datatable};
        \addlegendentry{$T_{int} = 0.12$};
        
        \addplot[black, solid, thick, mark=triangle, plot_style] 
            table [y=Genetic_Proposed_FDSA_SqExp_0_13_y, x=Genetic_Proposed_FDSA_SqExp_0_13_x, col sep=comma]{\datatable};
        \addlegendentry{$T_{int} = 0.13$};
        
        % \addplot[black, solid, thick, mark=hexagon, plot_style] 
        %     table [y=Genetic_Proposed_FDSA_SqExp_0_14_y, x=Genetic_Proposed_FDSA_SqExp_0_14_x, col sep=comma]{\datatable};
        % \addlegendentry{$T_{int} = 0.14$};
        
        \addplot[black, solid, thick, mark=o, plot_style] 
            table [y=Genetic_Proposed_FDSA_SqExp_0_15_y, x=Genetic_Proposed_FDSA_SqExp_0_15_x, col sep=comma]{\datatable};
        \addlegendentry{$T_{int} = 0.15$};
    \end{axis}
\end{tikzpicture}
        }
        \caption{Avg Frame Collision}
        \label{fig:data_GPD025_C05_S10PE_FC_TAU}
    \end{subfigure}%
    \begin{subfigure}{.33\textwidth}
        \resizebox{\linewidth}{!}{
            \pgfplotstableread[col sep = comma]{./data_new/data_GPD025_TP.csv}\datatable
            \pgfplotsset{ylabel=Avg. Throughput, ymin=3.75, ymax=4.35, ytick={3.75,3.85, ...,4.50}, ignore legend}
            \tikzstyle{plot_style} = [line width=1.25pt, mark size={4.0}, mark repeat=4, mark phase=1]
\begin{tikzpicture}[thick]
    \begin{axis}[
        width=8cm,
        height=7cm,
        xmin=0e3,
        xmax=12e3,
        grid=major,
        xlabel={Timesteps},
        xlabel style={font=\large},
        ylabel style={font=\large},
        ytick style={font=\tiny},
        yticklabel style={
            /pgf/number format/.cd,
                fixed,
                fixed zerofill,
                precision=2,
            /tikz/.cd
        },
        % log ticks with fixed point,
        % title={},
        legend pos=north east,
        legend cell align={left},
        legend style={fill opacity=0.80, draw opacity=0.50, text opacity=1.0}
        ]
        
        \addplot[black, solid, thick, mark=+, plot_style] 
            table [y=Genetic_Proposed_FDSA_SqExp_0_10_y, x=Genetic_Proposed_FDSA_SqExp_0_10_x, col sep=comma]{\datatable};
        \addlegendentry{$T_{int} = 0.10$};
        
        % \addplot[black, solid, thick, mark=triangle, plot_style] 
        %     table [y=Genetic_Proposed_FDSA_SqExp_0_11_y, x=Genetic_Proposed_FDSA_SqExp_0_11_x, col sep=comma]{\datatable};
        % \addlegendentry{$T_{int} = 0.11$};
        
        \addplot[black, solid, thick, mark=square, plot_style] 
            table [y=Genetic_Proposed_FDSA_SqExp_0_12_y, x=Genetic_Proposed_FDSA_SqExp_0_12_x, col sep=comma]{\datatable};
        \addlegendentry{$T_{int} = 0.12$};
        
        \addplot[black, solid, thick, mark=triangle, plot_style] 
            table [y=Genetic_Proposed_FDSA_SqExp_0_13_y, x=Genetic_Proposed_FDSA_SqExp_0_13_x, col sep=comma]{\datatable};
        \addlegendentry{$T_{int} = 0.13$};
        
        % \addplot[black, solid, thick, mark=hexagon, plot_style] 
        %     table [y=Genetic_Proposed_FDSA_SqExp_0_14_y, x=Genetic_Proposed_FDSA_SqExp_0_14_x, col sep=comma]{\datatable};
        % \addlegendentry{$T_{int} = 0.14$};
        
        \addplot[black, solid, thick, mark=o, plot_style] 
            table [y=Genetic_Proposed_FDSA_SqExp_0_15_y, x=Genetic_Proposed_FDSA_SqExp_0_15_x, col sep=comma]{\datatable};
        \addlegendentry{$T_{int} = 0.15$};
    \end{axis}
\end{tikzpicture}
        }
        \caption{Achievable Throughput}
        \label{fig:data_GPD025_C05_S10PE_TP_TAU}
    \end{subfigure}%
    \begin{subfigure}{.33\textwidth}
        \resizebox{\linewidth}{!}{
            \pgfplotstableread[col sep = comma]{./data_new/data_GPD025_SE.csv}\datatable
            \pgfplotsset{ylabel=Avg. sensing, ymin=0.70, ymax=1.00, ytick={0.70,0.75,...,1.05}}
            \tikzstyle{plot_style} = [line width=1.25pt, mark size={4.0}, mark repeat=4, mark phase=1]
\begin{tikzpicture}[thick]
    \begin{axis}[
        width=8cm,
        height=7cm,
        xmin=0e3,
        xmax=12e3,
        grid=major,
        xlabel={Timesteps},
        xlabel style={font=\large},
        ylabel style={font=\large},
        ytick style={font=\tiny},
        yticklabel style={
            /pgf/number format/.cd,
                fixed,
                fixed zerofill,
                precision=2,
            /tikz/.cd
        },
        % log ticks with fixed point,
        % title={},
        legend pos=north east,
        legend cell align={left},
        legend style={fill opacity=0.80, draw opacity=0.50, text opacity=1.0}
        ]
        
        \addplot[black, solid, thick, mark=+, plot_style] 
            table [y=Genetic_Proposed_FDSA_SqExp_0_10_y, x=Genetic_Proposed_FDSA_SqExp_0_10_x, col sep=comma]{\datatable};
        \addlegendentry{$T_{int} = 0.10$};
        
        % \addplot[black, solid, thick, mark=triangle, plot_style] 
        %     table [y=Genetic_Proposed_FDSA_SqExp_0_11_y, x=Genetic_Proposed_FDSA_SqExp_0_11_x, col sep=comma]{\datatable};
        % \addlegendentry{$T_{int} = 0.11$};
        
        \addplot[black, solid, thick, mark=square, plot_style] 
            table [y=Genetic_Proposed_FDSA_SqExp_0_12_y, x=Genetic_Proposed_FDSA_SqExp_0_12_x, col sep=comma]{\datatable};
        \addlegendentry{$T_{int} = 0.12$};
        
        \addplot[black, solid, thick, mark=triangle, plot_style] 
            table [y=Genetic_Proposed_FDSA_SqExp_0_13_y, x=Genetic_Proposed_FDSA_SqExp_0_13_x, col sep=comma]{\datatable};
        \addlegendentry{$T_{int} = 0.13$};
        
        % \addplot[black, solid, thick, mark=hexagon, plot_style] 
        %     table [y=Genetic_Proposed_FDSA_SqExp_0_14_y, x=Genetic_Proposed_FDSA_SqExp_0_14_x, col sep=comma]{\datatable};
        % \addlegendentry{$T_{int} = 0.14$};
        
        \addplot[black, solid, thick, mark=o, plot_style] 
            table [y=Genetic_Proposed_FDSA_SqExp_0_15_y, x=Genetic_Proposed_FDSA_SqExp_0_15_x, col sep=comma]{\datatable};
        \addlegendentry{$T_{int} = 0.15$};
    \end{axis}
\end{tikzpicture}
        }
        \caption{Avg No.of Sensing per active SU}
        \label{fig:data_GPD025_C05_S10PE_SE_TAU}
    \end{subfigure}%
    \caption{Effect of $T_{int}$ in GPD Model for $5$ channels and $10$ users using Proposed FDSA}
    \label{fig:GPD_Tint}
\end{figure*}
\fi

Traditionally, to simulate multiple users in CRNs, an assumption that the number of available primary channels is greater than the number of SUs is made \cite{gai2011combinatorial,liu2010distributed}. Now, in the era of increasing number of secondary devices, the above assumption does not hold true; we are dealing with more of devices than the number of available channels. Hence, we consider a scenario in which there are $5$ primary channels ($N=5$) and $10$ SU devices ($M=20$) that are competing for secondary access. It has been suggested through the study of real-life traces that heavy-tailed distributions like Generalized Pareto Distribution (GPD) are suited to model the distribution of the idle times of primary traffic  \cite{stabellini2010quantifying}. Also, in notable works like \cite{pei2007sensing}, the exponential distribution is used to capture the traffic behaviour of primary users. In the following sessions, we provide results in two different PU traffic models - GPD and Exponential. We model each channel independently where the ON times and idle times of the PU are independent and identically distributed (iid) samples from the respective distributions. The distribution for each channel is modelled with parameters randomly selected from the range mentioned in Table \ref{table:param}. \added{These traffic parameters are chosen to replicate a high traffic scenario as it is the most difficult case for opportunistic spectrum access. We also assume perfect sensing by the energy detector and perfect channel conditions to clearly demonstrate the effectiveness of the proposed method.}
% In order to make our simulations more realistic, we also account for the probability with which the SU's transmission might fail due to channel error. This implies that the failure in secondary transmission is not due to collision with the PU alone, a fraction of the failures is also due to channel error. Note that the central node cannot distinguish between these failures and hence treats all failed transmissions as packet collisions.
    
We consider SUs that transmit periodically once in $SU_{interval}$ frames for a duration of $SU_{ON}$ frames. The parameters are set such that there is a heavy demand for the primary channels in this case. Parameters used for the simulation are listed in Table \ref{table:param}. 
    
\begin{table}[!h]
    \caption{Parameters used for simulation}
    \label{table:param}    
    \centering
    \begin{tabular}{ |c|c| } 
         \hline
         \textbf{Parameter} & \textbf{Value} \\ 
         \hline
         \hline
         Continuous Traffic Model-GPD & $\sigma=25$, $k\in[0,0.1]$, $\theta\in[10,50]$ \\ 
        \hline
         Exponential Traffic Model & $\theta_{ON}^{-1}=[0,100], \theta_{OFF}^{-1}=[0,50]$ \\
        % \hline
        % HED Traffic Model & $\mu_i = 150$ \& $k = 4$  \\
        % \hline
        % $P_d$ & 0.95\\
        %  \hline
        % $P_f$ & 0.05\\
         \hline
         Frame duration ($T_f$) & $10ms$\\
         \hline
         Sensing duration ($\tau$) & $2ms$ \\   
        %  \hline
        %  SNR of SU at SU receiver & $20dB$ \\
        \hline
        SNR of PU at SU receiver & $-10dB$ \\
        \hline
        Number of Channels & $5$ \\
        \hline
        Number of Secondary Users & $10$ \\
        \hline
        $SU_{ON}$ for periodic devices & $5$ frames \\
        \hline
        $SU_{interval}$ for periodic devices & $100$ frames\\
        % \hline
        % $P_{alarm}$ for event driven devices & $0.05$\\
        % \hline 
        % ON Time for event driven devices & $\lambda^{-1} =10$\\
        \hline
        % Channel error & 0.05 \\
        % \hline
        % $E_{sense}$ & $62.5 \mu J$ \\
        % \hline
        % $E_{active}$ & $500 \mu J$ \\
        % \hline
        % $E_{wait}, E_{idle}$ & $5 \mu J$ \\
        % \hline
    \end{tabular}
\end{table}

We compare our method with the following approaches to benchmark the performance. 
\begin{enumerate}
    % \item \textbf{Genie}: This is the method which knows exact information about the state of PU channels and hence can select the best channel at each instant. This method assumed the knowledge of exact OFF time of primary users and vacates the primary channel when the PU is about to start transmission. Also this method assumes the channel quality at each PU channel for each SU and hence can compute the best SU-channel combination.
    \item \textbf{Transmit Until Collision (TUC)}: This is the classic method of transmit on the primary channel until a collision occurs (with PU) and then hop the channel. If the primary traffic if very low, this is one of the most effective strategies as the PU collisions will be less.
    \item \textbf{Traditional}: This method senses the PU channels before every packet transmission. Since sensing is performed at the start of every frame, the throughput is expected to be less with this method. However, this should also decrease the number of collisions.
    \item \textbf{\cite{raj2018spectrum}}: The method in \cite{raj2018spectrum} which uses a parametric method to predict the residual time is implemented.
    \item \textbf{Proposed with fixed $\bm{\epsilon}$}: In this variant, we use the proposed algorithm for both channel selection and transmission, but the exploration factor, $\epsilon$, is kept constant through out the process.
    \item \textbf{Proposed FDSA}: The proposed method using FDSA for adaptively modifying the exploration factor \(\epsilon\) to fit the specified threshold \(T_{int}\) is implemented. Note that the asymmetric cost function in (\ref{eqn:asymmetric}) is used.
\end{enumerate}
\added{All the comparative methods (Methods 1-3 listed above) focus on how often to sense for a single SU. We need to extend the algorithms to a multiple SU scenario so that they can be compared with the proposed algorithm. Hence, we append an adhoc channel selection algorithm, which randomly assigns the available channels to the requesting SUs, so as to cater to the multiple SU scenario.}
    
The parameter $\eta$, which corresponds to the fraction of times channel is selected at random instead of performing the hill climbing algorithm, is set to $0.2$. This is done to ensure that all the channels are sampled enough while building the value table. $\kappa$, set to $0.5$, denotes the rate at which the value table is built, i.e., the weight given to a newly observed sample in comparison with the previously maintained estimate for the value of the device-channel pairing, as specified in Algorithm \ref{alg:hub_channel}. The parameters for Algorithm \ref{alg:adaptive_exploration} determine the convergence of the descent algorithm; they are set as $a = 5, \alpha = 0.2, v = 0.1, \epsilon_0 = 1.0$ and  $\gamma = 0.4$ for all the simulations. $T_{int}$, the threshold for collision on each channel as seen by the SU, can be chosen based on a variety of factors such as the nature of primary traffic, the reliability and latency requirements of the SUs, etc. 

To quantify the performance of our algorithm, we consider the following metrics- throughput, the average number of frame collisions encountered and the number of sensing operations that are performed. As the SUs can achieve different maximum throughput on different channels, we model the capacity as a random number which is fixed for each user-channel pairing. All simulations are performed for a fixed set of values for the capacity. For all the metrics, i.e., throughput, number of sensing and frame collisions, we plot the cumulative values normalized to the number of frames the SU attempts to transmit till time $t$, say $F_t$ per active SU. Let $N_{active,t}$ refer to the number of SUs that are ON at a given time instant $t$. Then, for a metric $X$, we plot
\begin{align}
    y_t = \frac{1}{N_{active,t}} \frac{1}{F_t} \sum \limits_{n=1}^{t} X_{n}.
\end{align}
\added{All the metrics are averaged over \(200\) experiments.}

\ifCLASSOPTIONonecolumn
\begin{figure*}[t]
    \centering
    \begin{subfigure}{.33\textwidth}
        \resizebox{\linewidth}{!}{
            \pgfplotstableread[col sep = comma]{./data_new/data_Exp100050_FC.csv}\datatable
            \pgfplotsset{ylabel=Avg. frame collision, ymin=-0.05, ymax=+0.25, ytick={-0.05,0.0,...,+0.30}, ignore legend}
            \tikzstyle{plot_style} = [line width=1.25pt, mark size={4.0}, mark repeat=4, mark phase=1]
\begin{tikzpicture}[thick]
    \begin{axis}[
        width=8cm,
        height=7cm,
        xmin=0e3,
        xmax=14e3,
        grid=major,
        xlabel={Timesteps},
        xlabel style={font=\large},
        ylabel style={font=\large},
        ytick style={font=\tiny},
        yticklabel style={
            /pgf/number format/.cd,
                fixed,
                fixed zerofill,
                precision=2,
            /tikz/.cd
        },
        % log ticks with fixed point,
        % title={},
        % legend pos=
        legend style={at={(0.99,0.40)},anchor=south east},
        legend cell align={left},
        legend style={fill opacity=0.80, draw opacity=0.50, text opacity=1.0}
        ]
        
        % \addplot[black, solid, thick, mark=+, plot_style] 
        %     table [y=Genetic_TUC_y, x=Genetic_TUC_x, col sep=comma]{\datatable};
        % \addlegendentry{TUC};
        \addplot[black, solid, thick, mark=+, plot_style] 
            table [y=Random_TUC_y, x=Random_TUC_x, col sep=comma]{\datatable};
        \addlegendentry{TUC};
        
        % \addplot[black, solid, thick, mark=x, plot_style] 
        %     table [y=Genetic_Traditional_y, x=Genetic_Traditional_x, col sep=comma]{\datatable};
        % \addlegendentry{Traditional};
        \addplot[black, solid, thick, mark=x, plot_style] 
            table [y=Random_Traditional_y, x=Random_Traditional_x, col sep=comma]{\datatable};
        \addlegendentry{Traditional};
        
        % \addplot[black, solid, thick, mark=triangle, plot_style] 
        %     table [y=Genetic_Raj2018_y, x=Genetic_Raj2018_x, col sep=comma]{\datatable};
        % \addlegendentry{Augmented \cite{raj2018spectrum}};
        \addplot[black, solid, thick, mark=diamond, plot_style] 
            table [y=Random_Raj2018_y, x=Random_Raj2018_x, col sep=comma]{\datatable};
        \addlegendentry{\cite{raj2018spectrum}};
        
        % \addplot[black, solid, thick, mark=diamond, plot_style] 
        %     table [y=Genetic_Proposed_Fixed_eps_0_y, x=Genetic_Proposed_Fixed_eps_0_x, col sep=comma]{\datatable};
        % \addlegendentry{Proposed Greedy};
        
        \addplot[black, solid, thick, mark=square, plot_style] 
            table [y=Genetic_Proposed_Fixed_eps_0_1_y, x=Genetic_Proposed_Fixed_eps_0_1_x, col sep=comma]{\datatable};
        \addlegendentry{Proposed $(\epsilon = 0.10)$};
        
        % \addplot[black, solid, thick, mark=pentagon, plot_style] 
        %     table [y=Genetic_Proposed_Decreasing_eps_y, x=Genetic_Proposed_Decreasing_eps_x, col sep=comma]{\datatable};
        % \addlegendentry{Proposed ($\epsilon \downarrow$)};
        
        \addplot[black, solid, thick, mark=o, plot_style] 
            table [y=Genetic_Proposed_FDSA_SqExp_0_13_y, x=Genetic_Proposed_FDSA_SqExp_0_13_x, col sep=comma]{\datatable};
        \addlegendentry{Proposed FDSA $(T_{int} = 0.13)$};
    \end{axis}
\end{tikzpicture}
        }
        \caption{Avg Frame Collision}
        \label{fig:data_Exp100050_C05_S10PE_FC}
    \end{subfigure}%
    \begin{subfigure}{.33\textwidth}
        \resizebox{\linewidth}{!}{
            \pgfplotstableread[col sep = comma]{./data_new/data_Exp100050_TP.csv}\datatable
            \pgfplotsset{ylabel=Avg. Throughput, ymin=2.00, ymax=3.20, ytick={2.00,2.20,...,3.30}, ignore legend}
            \tikzstyle{plot_style} = [line width=1.25pt, mark size={4.0}, mark repeat=4, mark phase=1]
\begin{tikzpicture}[thick]
    \begin{axis}[
        width=8cm,
        height=7cm,
        xmin=0e3,
        xmax=14e3,
        grid=major,
        xlabel={Timesteps},
        xlabel style={font=\large},
        ylabel style={font=\large},
        ytick style={font=\tiny},
        yticklabel style={
            /pgf/number format/.cd,
                fixed,
                fixed zerofill,
                precision=2,
            /tikz/.cd
        },
        % log ticks with fixed point,
        % title={},
        % legend pos=
        legend style={at={(0.99,0.40)},anchor=south east},
        legend cell align={left},
        legend style={fill opacity=0.80, draw opacity=0.50, text opacity=1.0}
        ]
        
        % \addplot[black, solid, thick, mark=+, plot_style] 
        %     table [y=Genetic_TUC_y, x=Genetic_TUC_x, col sep=comma]{\datatable};
        % \addlegendentry{TUC};
        \addplot[black, solid, thick, mark=+, plot_style] 
            table [y=Random_TUC_y, x=Random_TUC_x, col sep=comma]{\datatable};
        \addlegendentry{TUC};
        
        % \addplot[black, solid, thick, mark=x, plot_style] 
        %     table [y=Genetic_Traditional_y, x=Genetic_Traditional_x, col sep=comma]{\datatable};
        % \addlegendentry{Traditional};
        \addplot[black, solid, thick, mark=x, plot_style] 
            table [y=Random_Traditional_y, x=Random_Traditional_x, col sep=comma]{\datatable};
        \addlegendentry{Traditional};
        
        % \addplot[black, solid, thick, mark=triangle, plot_style] 
        %     table [y=Genetic_Raj2018_y, x=Genetic_Raj2018_x, col sep=comma]{\datatable};
        % \addlegendentry{Augmented \cite{raj2018spectrum}};
        \addplot[black, solid, thick, mark=diamond, plot_style] 
            table [y=Random_Raj2018_y, x=Random_Raj2018_x, col sep=comma]{\datatable};
        \addlegendentry{\cite{raj2018spectrum}};
        
        % \addplot[black, solid, thick, mark=diamond, plot_style] 
        %     table [y=Genetic_Proposed_Fixed_eps_0_y, x=Genetic_Proposed_Fixed_eps_0_x, col sep=comma]{\datatable};
        % \addlegendentry{Proposed Greedy};
        
        \addplot[black, solid, thick, mark=square, plot_style] 
            table [y=Genetic_Proposed_Fixed_eps_0_1_y, x=Genetic_Proposed_Fixed_eps_0_1_x, col sep=comma]{\datatable};
        \addlegendentry{Proposed $(\epsilon = 0.10)$};
        
        % \addplot[black, solid, thick, mark=pentagon, plot_style] 
        %     table [y=Genetic_Proposed_Decreasing_eps_y, x=Genetic_Proposed_Decreasing_eps_x, col sep=comma]{\datatable};
        % \addlegendentry{Proposed ($\epsilon \downarrow$)};
        
        \addplot[black, solid, thick, mark=o, plot_style] 
            table [y=Genetic_Proposed_FDSA_SqExp_0_13_y, x=Genetic_Proposed_FDSA_SqExp_0_13_x, col sep=comma]{\datatable};
        \addlegendentry{Proposed FDSA $(T_{int} = 0.13)$};
    \end{axis}
\end{tikzpicture}
        }
        \caption{Achievable Throughput}
        \label{fig:data_Exp100050_C05_S10PE_TP}
    \end{subfigure}%
    \begin{subfigure}{.33\textwidth}
        \resizebox{\linewidth}{!}{
            \pgfplotstableread[col sep = comma]{./data_new/data_Exp100050_SE.csv}\datatable
            \pgfplotsset{ylabel=Avg. sensing, ymin=0.80,ymax=2.20, ytick={0.80,1.00,...,2.30}}
            \tikzstyle{plot_style} = [line width=1.25pt, mark size={4.0}, mark repeat=4, mark phase=1]
\begin{tikzpicture}[thick]
    \begin{axis}[
        width=8cm,
        height=7cm,
        xmin=0e3,
        xmax=14e3,
        grid=major,
        xlabel={Timesteps},
        xlabel style={font=\large},
        ylabel style={font=\large},
        ytick style={font=\tiny},
        yticklabel style={
            /pgf/number format/.cd,
                fixed,
                fixed zerofill,
                precision=2,
            /tikz/.cd
        },
        % log ticks with fixed point,
        % title={},
        % legend pos=
        legend style={at={(0.99,0.40)},anchor=south east},
        legend cell align={left},
        legend style={fill opacity=0.80, draw opacity=0.50, text opacity=1.0}
        ]
        
        % \addplot[black, solid, thick, mark=+, plot_style] 
        %     table [y=Genetic_TUC_y, x=Genetic_TUC_x, col sep=comma]{\datatable};
        % \addlegendentry{TUC};
        \addplot[black, solid, thick, mark=+, plot_style] 
            table [y=Random_TUC_y, x=Random_TUC_x, col sep=comma]{\datatable};
        \addlegendentry{TUC};
        
        % \addplot[black, solid, thick, mark=x, plot_style] 
        %     table [y=Genetic_Traditional_y, x=Genetic_Traditional_x, col sep=comma]{\datatable};
        % \addlegendentry{Traditional};
        \addplot[black, solid, thick, mark=x, plot_style] 
            table [y=Random_Traditional_y, x=Random_Traditional_x, col sep=comma]{\datatable};
        \addlegendentry{Traditional};
        
        % \addplot[black, solid, thick, mark=triangle, plot_style] 
        %     table [y=Genetic_Raj2018_y, x=Genetic_Raj2018_x, col sep=comma]{\datatable};
        % \addlegendentry{Augmented \cite{raj2018spectrum}};
        \addplot[black, solid, thick, mark=diamond, plot_style] 
            table [y=Random_Raj2018_y, x=Random_Raj2018_x, col sep=comma]{\datatable};
        \addlegendentry{\cite{raj2018spectrum}};
        
        % \addplot[black, solid, thick, mark=diamond, plot_style] 
        %     table [y=Genetic_Proposed_Fixed_eps_0_y, x=Genetic_Proposed_Fixed_eps_0_x, col sep=comma]{\datatable};
        % \addlegendentry{Proposed Greedy};
        
        \addplot[black, solid, thick, mark=square, plot_style] 
            table [y=Genetic_Proposed_Fixed_eps_0_1_y, x=Genetic_Proposed_Fixed_eps_0_1_x, col sep=comma]{\datatable};
        \addlegendentry{Proposed $(\epsilon = 0.10)$};
        
        % \addplot[black, solid, thick, mark=pentagon, plot_style] 
        %     table [y=Genetic_Proposed_Decreasing_eps_y, x=Genetic_Proposed_Decreasing_eps_x, col sep=comma]{\datatable};
        % \addlegendentry{Proposed ($\epsilon \downarrow$)};
        
        \addplot[black, solid, thick, mark=o, plot_style] 
            table [y=Genetic_Proposed_FDSA_SqExp_0_13_y, x=Genetic_Proposed_FDSA_SqExp_0_13_x, col sep=comma]{\datatable};
        \addlegendentry{Proposed FDSA $(T_{int} = 0.13)$};
    \end{axis}
\end{tikzpicture}
        }
        \caption{Avg No.of Sensing per active SU}
        \label{fig:data_Exp100050_C05_S10PE_SE}
    \end{subfigure}%
    \caption{Results for Exponential Model for $5$ channels and $10$ users for periodic SU traffic}
    \label{fig:Exp_comp}
\end{figure*}

\begin{figure*}[t]
    \centering
    \begin{subfigure}{.33\textwidth}
        \resizebox{\linewidth}{!}{
            \pgfplotstableread[col sep = comma]{./data_new/data_Exp100050_FC.csv}\datatable
            \pgfplotsset{ylabel=Avg. frame collision, ymin=0.11, ymax=0.17, ytick={0.11,0.12,...,0.18}, ignore legend}
            \tikzstyle{plot_style} = [line width=1.25pt, mark size={4.0}, mark repeat=4, mark phase=1]
\begin{tikzpicture}[thick]
    \begin{axis}[
        width=8cm,
        height=7cm,
        xmin=0e3,
        xmax=14e3,
        grid=major,
        xlabel={Timesteps},
        xlabel style={font=\large},
        ylabel style={font=\large},
        ytick style={font=\tiny},
        yticklabel style={
            /pgf/number format/.cd,
                fixed,
                fixed zerofill,
                precision=2,
            /tikz/.cd
        },
        % log ticks with fixed point,
        % title={},
        legend pos=north east,
        legend cell align={left},
        legend style={fill opacity=0.80, draw opacity=0.50, text opacity=1.0}
        ]
        
        \addplot[black, solid, thick, mark=triangle, plot_style] 
            table [y=Genetic_Proposed_FDSA_SqExp_0_11_y, x=Genetic_Proposed_FDSA_SqExp_0_11_x, col sep=comma]{\datatable};
        \addlegendentry{$T_{int} = 0.11$};
        
        \addplot[black, solid, thick, mark=square, plot_style] 
            table [y=Genetic_Proposed_FDSA_SqExp_0_13_y, x=Genetic_Proposed_FDSA_SqExp_0_13_x, col sep=comma]{\datatable};
        \addlegendentry{$T_{int} = 0.13$};
        
        \addplot[black, solid, thick, mark=pentagon, plot_style] 
            table [y=Genetic_Proposed_FDSA_SqExp_0_15_y, x=Genetic_Proposed_FDSA_SqExp_0_15_x, col sep=comma]{\datatable};
        \addlegendentry{$T_{int} = 0.15$};
        
        \addplot[black, solid, thick, mark=o, plot_style] 
            table [y=Genetic_Proposed_FDSA_SqExp_0_18_y, x=Genetic_Proposed_FDSA_SqExp_0_18_x, col sep=comma]{\datatable};
        \addlegendentry{$T_{int} = 0.18$};
    \end{axis}
\end{tikzpicture}
        }
        \caption{Avg Frame Collision}
        \label{fig:data_Exp100050_C05_S10PE_FC_TAU}
    \end{subfigure}%
    \begin{subfigure}{.33\textwidth}
        \resizebox{\linewidth}{!}{
            \pgfplotstableread[col sep = comma]{./data_new/data_Exp100050_TP.csv}\datatable
            \pgfplotsset{ylabel=Avg. Throughput, ymin=2.70, ymax=3.10, ytick={2.70, 2.80,...,3.20}, ignore legend}
            \tikzstyle{plot_style} = [line width=1.25pt, mark size={4.0}, mark repeat=4, mark phase=1]
\begin{tikzpicture}[thick]
    \begin{axis}[
        width=8cm,
        height=7cm,
        xmin=0e3,
        xmax=14e3,
        grid=major,
        xlabel={Timesteps},
        xlabel style={font=\large},
        ylabel style={font=\large},
        ytick style={font=\tiny},
        yticklabel style={
            /pgf/number format/.cd,
                fixed,
                fixed zerofill,
                precision=2,
            /tikz/.cd
        },
        % log ticks with fixed point,
        % title={},
        legend pos=north east,
        legend cell align={left},
        legend style={fill opacity=0.80, draw opacity=0.50, text opacity=1.0}
        ]
        
        \addplot[black, solid, thick, mark=triangle, plot_style] 
            table [y=Genetic_Proposed_FDSA_SqExp_0_11_y, x=Genetic_Proposed_FDSA_SqExp_0_11_x, col sep=comma]{\datatable};
        \addlegendentry{$T_{int} = 0.11$};
        
        \addplot[black, solid, thick, mark=square, plot_style] 
            table [y=Genetic_Proposed_FDSA_SqExp_0_13_y, x=Genetic_Proposed_FDSA_SqExp_0_13_x, col sep=comma]{\datatable};
        \addlegendentry{$T_{int} = 0.13$};
        
        \addplot[black, solid, thick, mark=pentagon, plot_style] 
            table [y=Genetic_Proposed_FDSA_SqExp_0_15_y, x=Genetic_Proposed_FDSA_SqExp_0_15_x, col sep=comma]{\datatable};
        \addlegendentry{$T_{int} = 0.15$};
        
        \addplot[black, solid, thick, mark=o, plot_style] 
            table [y=Genetic_Proposed_FDSA_SqExp_0_18_y, x=Genetic_Proposed_FDSA_SqExp_0_18_x, col sep=comma]{\datatable};
        \addlegendentry{$T_{int} = 0.18$};
    \end{axis}
\end{tikzpicture}
        }
        \caption{Achievable Throughput}
        \label{fig:data_Exp100050_C05_S10PE_TP_TAU}
    \end{subfigure}%
    \begin{subfigure}{.33\textwidth}
        \resizebox{\linewidth}{!}{
            \pgfplotstableread[col sep = comma]{./data_new/data_Exp100050_SE.csv}\datatable
            \pgfplotsset{ylabel=Avg. sensing, ymin=0.80, ymax=1.40, ytick={0.90,1.00,...,1.30}}
            \tikzstyle{plot_style} = [line width=1.25pt, mark size={4.0}, mark repeat=4, mark phase=1]
\begin{tikzpicture}[thick]
    \begin{axis}[
        width=8cm,
        height=7cm,
        xmin=0e3,
        xmax=14e3,
        grid=major,
        xlabel={Timesteps},
        xlabel style={font=\large},
        ylabel style={font=\large},
        ytick style={font=\tiny},
        yticklabel style={
            /pgf/number format/.cd,
                fixed,
                fixed zerofill,
                precision=2,
            /tikz/.cd
        },
        % log ticks with fixed point,
        % title={},
        legend pos=north east,
        legend cell align={left},
        legend style={fill opacity=0.80, draw opacity=0.50, text opacity=1.0}
        ]
        
        \addplot[black, solid, thick, mark=triangle, plot_style] 
            table [y=Genetic_Proposed_FDSA_SqExp_0_11_y, x=Genetic_Proposed_FDSA_SqExp_0_11_x, col sep=comma]{\datatable};
        \addlegendentry{$T_{int} = 0.11$};
        
        \addplot[black, solid, thick, mark=square, plot_style] 
            table [y=Genetic_Proposed_FDSA_SqExp_0_13_y, x=Genetic_Proposed_FDSA_SqExp_0_13_x, col sep=comma]{\datatable};
        \addlegendentry{$T_{int} = 0.13$};
        
        \addplot[black, solid, thick, mark=pentagon, plot_style] 
            table [y=Genetic_Proposed_FDSA_SqExp_0_15_y, x=Genetic_Proposed_FDSA_SqExp_0_15_x, col sep=comma]{\datatable};
        \addlegendentry{$T_{int} = 0.15$};
        
        \addplot[black, solid, thick, mark=o, plot_style] 
            table [y=Genetic_Proposed_FDSA_SqExp_0_18_y, x=Genetic_Proposed_FDSA_SqExp_0_18_x, col sep=comma]{\datatable};
        \addlegendentry{$T_{int} = 0.18$};
    \end{axis}
\end{tikzpicture}
        }
        \caption{Avg No.of Sensing per active SU}
        \label{fig:data_Exp100050_C05_S10PE_SE_TAU}
    \end{subfigure}%
    \caption{Effect of $T_{int}$ for Exponential Model for $5$ channels and $10$ users using Proposed FDSA}
    \label{fig:exp_Tint}
\end{figure*}
\fi

\added{
The comparison of the proposed method against other benchmarks in an exponential traffic model is shown in Fig. \ref{fig:Exp_comp}. We can see that the proposed methods with both fixed \(\epsilon\) as well as an adaptive \(\epsilon \) as determined by FDSA are able to achieve a fair trade-off between frame collisions and sensing operations. We also note that the throughput of the proposed methods are greater than the others. We can attribute this difference to the proposed channel assignment algorithm; the low throughput of the traditional method can be attributed to the use of a significant fraction of every frame duration for sensing. Although \cite{raj2018spectrum} results in the lowest sensing operations, we can observe that the number of collisions it causes to the primary user is very high, which is highly undesirable.

To further demonstrate the performance of our method with respect to adherence of the interference threshold, \(T_{int}\), we provide results in Fig. \ref{fig:exp_Tint}. We can clearly observe that the proposed method is able to adapt to the different thresholds provided. The proposed method that uses FDSA for adaptively changing \(\epsilon\) always converges to a value that is lesser than the provided threshold. This ensures the maximum throughput and the minimum number of sensing operations, which is ideal from the secondary users' perspective. For higher thresholds like \(T_{int}=0.18\), it is noted that although the frame collision does not reach \(T_{int}\), it still does not violate the threshold. This is because of the asymmetric loss function employed to ensure that the collision does not cross the threshold; the exploration factor \(\epsilon\) is rapidly decreased if it crosses the threshold even at one instance. }
% As the proposed method is designed more towards the goal of reducing the PU frame collisions (through adaptive exploration factor), it can be seen that the cost is paid in terms of throughput. The proposed method is able to provide a better frame collision and sensing performance. In case of scenarios like IoT devices which need low throughput but energy efficient transmissions, this can be an attractive factor. In Fig. \ref{fig:EXP150150_C05_S10PE}, the performance on a less busy channel (mean ON and OFF time sampled uniformly from (0,150]) PU channel is provided. We can observe that the proposed method is able to provide all good qualities of Transmit Until Collision, but with a lesser rate of frame collisions.

\ifCLASSOPTIONonecolumn
\begin{figure*}[t]
    \centering
    \begin{subfigure}{.33\textwidth}
        \resizebox{\linewidth}{!}{
            \pgfplotstableread[col sep = comma]{./data_new/data_GPD025_FC.csv}\datatable
            \pgfplotsset{ylabel=Avg. Frame collision, ymin=-0.05, ymax=+0.20, ytick={-0.05,0.00,...,+0.25}, ignore legend}
            \tikzstyle{plot_style} = [line width=1.25pt, mark size={4.0}, mark repeat=4, mark phase=1]
\begin{tikzpicture}[thick]
    \begin{axis}[
        width=8cm,
        height=7cm,
        xmin=0e3,
        xmax=12e3,
        grid=major,
        xlabel={Timesteps},
        xlabel style={font=\large},
        ylabel style={font=\large},
        ytick style={font=\tiny},
        yticklabel style={
            /pgf/number format/.cd,
                fixed,
                fixed zerofill,
                precision=2,
            /tikz/.cd
        },
        % log ticks with fixed point,
        % title={},
        % legend pos=south east,
        legend style={at={(0.99,0.40)},anchor=south east},
        legend cell align={left},
        legend style={fill opacity=0.80, draw opacity=0.50, text opacity=1.0}
        ]
        
        % \addplot[black, solid, thick, mark=+, plot_style] 
        %     table [y=Genetic_TUC_y, x=Genetic_TUC_x, col sep=comma]{\datatable};
        % \addlegendentry{TUC};
        \addplot[black, solid, thick, mark=+, plot_style] 
            table [y=Random_TUC_y, x=Random_TUC_x, col sep=comma]{\datatable};
        \addlegendentry{TUC};
        
        % \addplot[black, solid, thick, mark=x, plot_style] 
        %     table [y=Genetic_Traditional_y, x=Genetic_Traditional_x, col sep=comma]{\datatable};
        % \addlegendentry{Traditional};
        \addplot[black, solid, thick, mark=x, plot_style] 
            table [y=Random_Traditional_y, x=Random_Traditional_x, col sep=comma]{\datatable};
        \addlegendentry{Traditional};
        
        % \addplot[black, solid, thick, mark=triangle, plot_style] 
        %     table [y=Genetic_Raj2018_y, x=Genetic_Raj2018_x, col sep=comma]{\datatable};
        % \addlegendentry{Augmented \cite{raj2018spectrum}};
        \addplot[black, solid, thick, mark=diamond, plot_style] 
            table [y=Random_Raj2018_y, x=Random_Raj2018_x, col sep=comma]{\datatable};
        \addlegendentry{\cite{raj2018spectrum}};
        
        % \addplot[blue, solid, thick, mark=diamond, plot_style] 
        %     table [y=Random2_Raj2018_y, x=Random2_Raj2018_x, col sep=comma]{\datatable};
        % \addlegendentry{Adhoc2 \cite{raj2018spectrum}}
        
        % \addplot[black, solid, thick, mark=diamond, plot_style] 
        %     table [y=Genetic_Proposed_Fixed_eps_0_y, x=Genetic_Proposed_Fixed_eps_0_x, col sep=comma]{\datatable};
        % \addlegendentry{Proposed Greedy};
        
        \addplot[black, solid, thick, mark=square, plot_style] 
            table [y=Genetic_Proposed_Fixed_eps_0_15_y, x=Genetic_Proposed_Fixed_eps_0_15_x, col sep=comma]{\datatable};
        \addlegendentry{Proposed $(\epsilon = 0.15)$};
        
        % \addplot[black, solid, thick, mark=pentagon, plot_style] 
        %     table [y=Genetic_Proposed_Decreasing_eps_y, x=Genetic_Proposed_Decreasing_eps_x, col sep=comma]{\datatable};
        % \addlegendentry{Proposed ($\epsilon \downarrow$)};
        
        \addplot[black, solid, thick, mark=o, plot_style] 
            table [y=Genetic_Proposed_FDSA_SqExp_0_13_y, x=Genetic_Proposed_FDSA_SqExp_0_13_x, col sep=comma]{\datatable};
        \addlegendentry{Proposed FDSA $(T_{int}=0.13)$};
    \end{axis}
\end{tikzpicture}
        }
        \caption{Avg Frame Collision}
        \label{fig:data_GPD025_C05_S10PE_FC}
    \end{subfigure}%
    \begin{subfigure}{.33\textwidth}
        \resizebox{\linewidth}{!}{
            \pgfplotstableread[col sep = comma]{./data_new/data_GPD025_TP.csv}\datatable
            \pgfplotsset{ylabel=Avg. Throughput, ymin=2.5, ymax=4.50, ytick={2.50,3.00,...,4.50}, ignore legend}
            \tikzstyle{plot_style} = [line width=1.25pt, mark size={4.0}, mark repeat=4, mark phase=1]
\begin{tikzpicture}[thick]
    \begin{axis}[
        width=8cm,
        height=7cm,
        xmin=0e3,
        xmax=12e3,
        grid=major,
        xlabel={Timesteps},
        xlabel style={font=\large},
        ylabel style={font=\large},
        ytick style={font=\tiny},
        yticklabel style={
            /pgf/number format/.cd,
                fixed,
                fixed zerofill,
                precision=2,
            /tikz/.cd
        },
        % log ticks with fixed point,
        % title={},
        % legend pos=south east,
        legend style={at={(0.99,0.40)},anchor=south east},
        legend cell align={left},
        legend style={fill opacity=0.80, draw opacity=0.50, text opacity=1.0}
        ]
        
        % \addplot[black, solid, thick, mark=+, plot_style] 
        %     table [y=Genetic_TUC_y, x=Genetic_TUC_x, col sep=comma]{\datatable};
        % \addlegendentry{TUC};
        \addplot[black, solid, thick, mark=+, plot_style] 
            table [y=Random_TUC_y, x=Random_TUC_x, col sep=comma]{\datatable};
        \addlegendentry{TUC};
        
        % \addplot[black, solid, thick, mark=x, plot_style] 
        %     table [y=Genetic_Traditional_y, x=Genetic_Traditional_x, col sep=comma]{\datatable};
        % \addlegendentry{Traditional};
        \addplot[black, solid, thick, mark=x, plot_style] 
            table [y=Random_Traditional_y, x=Random_Traditional_x, col sep=comma]{\datatable};
        \addlegendentry{Traditional};
        
        % \addplot[black, solid, thick, mark=triangle, plot_style] 
        %     table [y=Genetic_Raj2018_y, x=Genetic_Raj2018_x, col sep=comma]{\datatable};
        % \addlegendentry{Augmented \cite{raj2018spectrum}};
        \addplot[black, solid, thick, mark=diamond, plot_style] 
            table [y=Random_Raj2018_y, x=Random_Raj2018_x, col sep=comma]{\datatable};
        \addlegendentry{\cite{raj2018spectrum}};
        
        % \addplot[blue, solid, thick, mark=diamond, plot_style] 
        %     table [y=Random2_Raj2018_y, x=Random2_Raj2018_x, col sep=comma]{\datatable};
        % \addlegendentry{Adhoc2 \cite{raj2018spectrum}}
        
        % \addplot[black, solid, thick, mark=diamond, plot_style] 
        %     table [y=Genetic_Proposed_Fixed_eps_0_y, x=Genetic_Proposed_Fixed_eps_0_x, col sep=comma]{\datatable};
        % \addlegendentry{Proposed Greedy};
        
        \addplot[black, solid, thick, mark=square, plot_style] 
            table [y=Genetic_Proposed_Fixed_eps_0_15_y, x=Genetic_Proposed_Fixed_eps_0_15_x, col sep=comma]{\datatable};
        \addlegendentry{Proposed $(\epsilon = 0.15)$};
        
        % \addplot[black, solid, thick, mark=pentagon, plot_style] 
        %     table [y=Genetic_Proposed_Decreasing_eps_y, x=Genetic_Proposed_Decreasing_eps_x, col sep=comma]{\datatable};
        % \addlegendentry{Proposed ($\epsilon \downarrow$)};
        
        \addplot[black, solid, thick, mark=o, plot_style] 
            table [y=Genetic_Proposed_FDSA_SqExp_0_13_y, x=Genetic_Proposed_FDSA_SqExp_0_13_x, col sep=comma]{\datatable};
        \addlegendentry{Proposed FDSA $(T_{int}=0.13)$};
    \end{axis}
\end{tikzpicture}
        }
        \caption{Achievable Throughput}
        \label{fig:data_GPD025_C05_S10PE_TP}
    \end{subfigure}%
    \begin{subfigure}{.33\textwidth}
        \resizebox{\linewidth}{!}{
            \pgfplotstableread[col sep = comma]{./data_new/data_GPD025_SE.csv}\datatable
            \pgfplotsset{ylabel=Avg. sensing, ymin=0.60, ymax=1.80, ytick={0.60,0.80,...,1.80}}
            \tikzstyle{plot_style} = [line width=1.25pt, mark size={4.0}, mark repeat=4, mark phase=1]
\begin{tikzpicture}[thick]
    \begin{axis}[
        width=8cm,
        height=7cm,
        xmin=0e3,
        xmax=12e3,
        grid=major,
        xlabel={Timesteps},
        xlabel style={font=\large},
        ylabel style={font=\large},
        ytick style={font=\tiny},
        yticklabel style={
            /pgf/number format/.cd,
                fixed,
                fixed zerofill,
                precision=2,
            /tikz/.cd
        },
        % log ticks with fixed point,
        % title={},
        % legend pos=south east,
        legend style={at={(0.99,0.40)},anchor=south east},
        legend cell align={left},
        legend style={fill opacity=0.80, draw opacity=0.50, text opacity=1.0}
        ]
        
        % \addplot[black, solid, thick, mark=+, plot_style] 
        %     table [y=Genetic_TUC_y, x=Genetic_TUC_x, col sep=comma]{\datatable};
        % \addlegendentry{TUC};
        \addplot[black, solid, thick, mark=+, plot_style] 
            table [y=Random_TUC_y, x=Random_TUC_x, col sep=comma]{\datatable};
        \addlegendentry{TUC};
        
        % \addplot[black, solid, thick, mark=x, plot_style] 
        %     table [y=Genetic_Traditional_y, x=Genetic_Traditional_x, col sep=comma]{\datatable};
        % \addlegendentry{Traditional};
        \addplot[black, solid, thick, mark=x, plot_style] 
            table [y=Random_Traditional_y, x=Random_Traditional_x, col sep=comma]{\datatable};
        \addlegendentry{Traditional};
        
        % \addplot[black, solid, thick, mark=triangle, plot_style] 
        %     table [y=Genetic_Raj2018_y, x=Genetic_Raj2018_x, col sep=comma]{\datatable};
        % \addlegendentry{Augmented \cite{raj2018spectrum}};
        \addplot[black, solid, thick, mark=diamond, plot_style] 
            table [y=Random_Raj2018_y, x=Random_Raj2018_x, col sep=comma]{\datatable};
        \addlegendentry{\cite{raj2018spectrum}};
        
        % \addplot[blue, solid, thick, mark=diamond, plot_style] 
        %     table [y=Random2_Raj2018_y, x=Random2_Raj2018_x, col sep=comma]{\datatable};
        % \addlegendentry{Adhoc2 \cite{raj2018spectrum}}
        
        % \addplot[black, solid, thick, mark=diamond, plot_style] 
        %     table [y=Genetic_Proposed_Fixed_eps_0_y, x=Genetic_Proposed_Fixed_eps_0_x, col sep=comma]{\datatable};
        % \addlegendentry{Proposed Greedy};
        
        \addplot[black, solid, thick, mark=square, plot_style] 
            table [y=Genetic_Proposed_Fixed_eps_0_15_y, x=Genetic_Proposed_Fixed_eps_0_15_x, col sep=comma]{\datatable};
        \addlegendentry{Proposed $(\epsilon = 0.15)$};
        
        % \addplot[black, solid, thick, mark=pentagon, plot_style] 
        %     table [y=Genetic_Proposed_Decreasing_eps_y, x=Genetic_Proposed_Decreasing_eps_x, col sep=comma]{\datatable};
        % \addlegendentry{Proposed ($\epsilon \downarrow$)};
        
        \addplot[black, solid, thick, mark=o, plot_style] 
            table [y=Genetic_Proposed_FDSA_SqExp_0_13_y, x=Genetic_Proposed_FDSA_SqExp_0_13_x, col sep=comma]{\datatable};
        \addlegendentry{Proposed FDSA $(T_{int}=0.13)$};
    \end{axis}
\end{tikzpicture}
        }
        \caption{Avg No.of Sensing per active SU}
        \label{fig:data_GPD025_C05_S10PE_SE}
    \end{subfigure}%
    \caption{Results for GPD Model for $5$ channels and $10$ users for periodic SU traffic}
    \label{fig:GPD_comp}
\end{figure*}

\begin{figure*}[t]
    \centering
    \begin{subfigure}{.33\textwidth}
        \resizebox{\linewidth}{!}{
            \pgfplotstableread[col sep = comma]{./data_new/data_GPD025_FC.csv}\datatable
            \pgfplotsset{ylabel=Avg. Frame collision, ymin=0.09, ymax=0.14, ignore legend}
            \tikzstyle{plot_style} = [line width=1.25pt, mark size={4.0}, mark repeat=4, mark phase=1]
\begin{tikzpicture}[thick]
    \begin{axis}[
        width=8cm,
        height=7cm,
        xmin=0e3,
        xmax=12e3,
        grid=major,
        xlabel={Timesteps},
        xlabel style={font=\large},
        ylabel style={font=\large},
        ytick style={font=\tiny},
        yticklabel style={
            /pgf/number format/.cd,
                fixed,
                fixed zerofill,
                precision=2,
            /tikz/.cd
        },
        % log ticks with fixed point,
        % title={},
        legend pos=north east,
        legend cell align={left},
        legend style={fill opacity=0.80, draw opacity=0.50, text opacity=1.0}
        ]
        
        \addplot[black, solid, thick, mark=+, plot_style] 
            table [y=Genetic_Proposed_FDSA_SqExp_0_10_y, x=Genetic_Proposed_FDSA_SqExp_0_10_x, col sep=comma]{\datatable};
        \addlegendentry{$T_{int} = 0.10$};
        
        % \addplot[black, solid, thick, mark=triangle, plot_style] 
        %     table [y=Genetic_Proposed_FDSA_SqExp_0_11_y, x=Genetic_Proposed_FDSA_SqExp_0_11_x, col sep=comma]{\datatable};
        % \addlegendentry{$T_{int} = 0.11$};
        
        \addplot[black, solid, thick, mark=square, plot_style] 
            table [y=Genetic_Proposed_FDSA_SqExp_0_12_y, x=Genetic_Proposed_FDSA_SqExp_0_12_x, col sep=comma]{\datatable};
        \addlegendentry{$T_{int} = 0.12$};
        
        \addplot[black, solid, thick, mark=triangle, plot_style] 
            table [y=Genetic_Proposed_FDSA_SqExp_0_13_y, x=Genetic_Proposed_FDSA_SqExp_0_13_x, col sep=comma]{\datatable};
        \addlegendentry{$T_{int} = 0.13$};
        
        % \addplot[black, solid, thick, mark=hexagon, plot_style] 
        %     table [y=Genetic_Proposed_FDSA_SqExp_0_14_y, x=Genetic_Proposed_FDSA_SqExp_0_14_x, col sep=comma]{\datatable};
        % \addlegendentry{$T_{int} = 0.14$};
        
        \addplot[black, solid, thick, mark=o, plot_style] 
            table [y=Genetic_Proposed_FDSA_SqExp_0_15_y, x=Genetic_Proposed_FDSA_SqExp_0_15_x, col sep=comma]{\datatable};
        \addlegendentry{$T_{int} = 0.15$};
    \end{axis}
\end{tikzpicture}
        }
        \caption{Avg Frame Collision}
        \label{fig:data_GPD025_C05_S10PE_FC_TAU}
    \end{subfigure}%
    \begin{subfigure}{.33\textwidth}
        \resizebox{\linewidth}{!}{
            \pgfplotstableread[col sep = comma]{./data_new/data_GPD025_TP.csv}\datatable
            \pgfplotsset{ylabel=Avg. Throughput, ymin=3.75, ymax=4.35, ytick={3.75,3.85, ...,4.50}, ignore legend}
            \tikzstyle{plot_style} = [line width=1.25pt, mark size={4.0}, mark repeat=4, mark phase=1]
\begin{tikzpicture}[thick]
    \begin{axis}[
        width=8cm,
        height=7cm,
        xmin=0e3,
        xmax=12e3,
        grid=major,
        xlabel={Timesteps},
        xlabel style={font=\large},
        ylabel style={font=\large},
        ytick style={font=\tiny},
        yticklabel style={
            /pgf/number format/.cd,
                fixed,
                fixed zerofill,
                precision=2,
            /tikz/.cd
        },
        % log ticks with fixed point,
        % title={},
        legend pos=north east,
        legend cell align={left},
        legend style={fill opacity=0.80, draw opacity=0.50, text opacity=1.0}
        ]
        
        \addplot[black, solid, thick, mark=+, plot_style] 
            table [y=Genetic_Proposed_FDSA_SqExp_0_10_y, x=Genetic_Proposed_FDSA_SqExp_0_10_x, col sep=comma]{\datatable};
        \addlegendentry{$T_{int} = 0.10$};
        
        % \addplot[black, solid, thick, mark=triangle, plot_style] 
        %     table [y=Genetic_Proposed_FDSA_SqExp_0_11_y, x=Genetic_Proposed_FDSA_SqExp_0_11_x, col sep=comma]{\datatable};
        % \addlegendentry{$T_{int} = 0.11$};
        
        \addplot[black, solid, thick, mark=square, plot_style] 
            table [y=Genetic_Proposed_FDSA_SqExp_0_12_y, x=Genetic_Proposed_FDSA_SqExp_0_12_x, col sep=comma]{\datatable};
        \addlegendentry{$T_{int} = 0.12$};
        
        \addplot[black, solid, thick, mark=triangle, plot_style] 
            table [y=Genetic_Proposed_FDSA_SqExp_0_13_y, x=Genetic_Proposed_FDSA_SqExp_0_13_x, col sep=comma]{\datatable};
        \addlegendentry{$T_{int} = 0.13$};
        
        % \addplot[black, solid, thick, mark=hexagon, plot_style] 
        %     table [y=Genetic_Proposed_FDSA_SqExp_0_14_y, x=Genetic_Proposed_FDSA_SqExp_0_14_x, col sep=comma]{\datatable};
        % \addlegendentry{$T_{int} = 0.14$};
        
        \addplot[black, solid, thick, mark=o, plot_style] 
            table [y=Genetic_Proposed_FDSA_SqExp_0_15_y, x=Genetic_Proposed_FDSA_SqExp_0_15_x, col sep=comma]{\datatable};
        \addlegendentry{$T_{int} = 0.15$};
    \end{axis}
\end{tikzpicture}
        }
        \caption{Achievable Throughput}
        \label{fig:data_GPD025_C05_S10PE_TP_TAU}
    \end{subfigure}%
    \begin{subfigure}{.33\textwidth}
        \resizebox{\linewidth}{!}{
            \pgfplotstableread[col sep = comma]{./data_new/data_GPD025_SE.csv}\datatable
            \pgfplotsset{ylabel=Avg. sensing, ymin=0.70, ymax=1.00, ytick={0.70,0.75,...,1.05}}
            \tikzstyle{plot_style} = [line width=1.25pt, mark size={4.0}, mark repeat=4, mark phase=1]
\begin{tikzpicture}[thick]
    \begin{axis}[
        width=8cm,
        height=7cm,
        xmin=0e3,
        xmax=12e3,
        grid=major,
        xlabel={Timesteps},
        xlabel style={font=\large},
        ylabel style={font=\large},
        ytick style={font=\tiny},
        yticklabel style={
            /pgf/number format/.cd,
                fixed,
                fixed zerofill,
                precision=2,
            /tikz/.cd
        },
        % log ticks with fixed point,
        % title={},
        legend pos=north east,
        legend cell align={left},
        legend style={fill opacity=0.80, draw opacity=0.50, text opacity=1.0}
        ]
        
        \addplot[black, solid, thick, mark=+, plot_style] 
            table [y=Genetic_Proposed_FDSA_SqExp_0_10_y, x=Genetic_Proposed_FDSA_SqExp_0_10_x, col sep=comma]{\datatable};
        \addlegendentry{$T_{int} = 0.10$};
        
        % \addplot[black, solid, thick, mark=triangle, plot_style] 
        %     table [y=Genetic_Proposed_FDSA_SqExp_0_11_y, x=Genetic_Proposed_FDSA_SqExp_0_11_x, col sep=comma]{\datatable};
        % \addlegendentry{$T_{int} = 0.11$};
        
        \addplot[black, solid, thick, mark=square, plot_style] 
            table [y=Genetic_Proposed_FDSA_SqExp_0_12_y, x=Genetic_Proposed_FDSA_SqExp_0_12_x, col sep=comma]{\datatable};
        \addlegendentry{$T_{int} = 0.12$};
        
        \addplot[black, solid, thick, mark=triangle, plot_style] 
            table [y=Genetic_Proposed_FDSA_SqExp_0_13_y, x=Genetic_Proposed_FDSA_SqExp_0_13_x, col sep=comma]{\datatable};
        \addlegendentry{$T_{int} = 0.13$};
        
        % \addplot[black, solid, thick, mark=hexagon, plot_style] 
        %     table [y=Genetic_Proposed_FDSA_SqExp_0_14_y, x=Genetic_Proposed_FDSA_SqExp_0_14_x, col sep=comma]{\datatable};
        % \addlegendentry{$T_{int} = 0.14$};
        
        \addplot[black, solid, thick, mark=o, plot_style] 
            table [y=Genetic_Proposed_FDSA_SqExp_0_15_y, x=Genetic_Proposed_FDSA_SqExp_0_15_x, col sep=comma]{\datatable};
        \addlegendentry{$T_{int} = 0.15$};
    \end{axis}
\end{tikzpicture}
        }
        \caption{Avg No.of Sensing per active SU}
        \label{fig:data_GPD025_C05_S10PE_SE_TAU}
    \end{subfigure}%
    \caption{Effect of $T_{int}$ in GPD Model for $5$ channels and $10$ users using Proposed FDSA}
    \label{fig:GPD_Tint}
\end{figure*}
\fi

\added{In Fig. \ref{fig:GPD_comp}, the results for the proposed algorithm in GPD primary traffic are shown in comparison with the other benchmarks. The trend observed is similar to the one observed in exponential traffic. The effectiveness of our channel assignment algorithm is exhibited through the improvement in throughput and the residual prediction algorithm in the number of sensings. We demonstrate the utility of our adaptive exploration algorithm through Fig. \ref{fig:GPD_Tint} where the algorithm adapts its exploration according to the prescribed collision threshold. We believe that such a method deeply benefits the SU to maximally exploit the available spectrum while adhering to the interference constraints set by the licensed user. }

% The results of HED traffic model is provided in Fig. \ref{fig:HED150150_C05_S10PE}. Since HED is also a low traffic model, all algorithms can achieve low collision rate. Also, both the variants of the proposed algorithm are able to give sensing performance almost close to that of genie. Here again, the cost is paid in terms of the decreased throughput.

% From all the simulations, we can observe that the proposed algorithm is able to achieve what it is designed for - achieve lower frame collisions and minimise the number of sensing required for transmission. As a trade-off, the proposed algorithm scarifies high throughput for more reliable and energy efficient transmission.
    
    \section{Concluding Remarks} \label{sec:Conc}
In this work, a centralized multi-stage algorithm was proposed for a cognitive radio network consisting of multiple SUs. Firstly, to assign an SU to a suitable channel, a hill climbing approach was advocated to lessen the computational burden of conducting a brute force search through all SU-channel combinations. 
% Once a channel was assigned, a skipping method was used to reduce sensing by the SU. This was achieved by learning the primary user statistic through a non-parametric learning method and using the learnt OFF time to skip sensing. 
Once a channel was assigned, a non-parametric Bayesian learning method using the Dirichlet prior was employed to estimate the distribution of the residual primary OFF time, which in turn was used to predict the number of frames for which one can skip sensing the channel. Further, to leverage a given threshold for collision, we adaptively trade-off transmitting until collision and choosing the time to skip from the learnt OFF time distribution by employing a stochastic approximation method, FDSA. We show through exhaustive simulations that the proposed method requires significantly lesser number of channel sensings and achieves comparable throughput while adhering to the collision threshold imposed when compared to the method which senses every frame. In an energy constrained scenario, this helps in improving the energy efficiency of the SUs without sacrificing spectrum efficiency. 

In this work, as we focus on the scenario where the number of secondary devices connecting to the network is greater than the number of channels, we advocate a centralized structure where the central node is responsible for coordination. 
% As the IoT ecosystem grows, we will see more resource constrained devices getting into the network. Cognitive capabilities should be built into these networks, either into the devices themselves or as a central entity to respond to the rapidly evolving requirements of the heterogeneous collection of devices. In this paper, we presented a centralized learning algorithm where the intelligence is embedded in a central entity.
Application of distributed learning techniques which are energy efficient could further help the devices to be more autonomous and the network to be more flexible for networks with lesser secondary devices. We believe that further effort towards the development of energy efficient AI/RL techniques for edge devices can substantially contribute to the improvement of CRNs.
	
	\bibliographystyle{IEEEtran}
	\bibliography{IEEEabrv,tccn.bib}
\end{document}